\documentclass[preprint,onecolumn,superscriptaddress,nofootinbib,11pt]{article}
\pdfoutput=1
\usepackage{graphicx,epsfig,enumitem,amsmath,latexsym,amssymb,hyperref,enumerate,amsfonts,mathrsfs,url,color,cancel,cite,xcolor,feynmp,slashed,verbatim}
\usepackage[utf8]{inputenc}
\usepackage[normalem]{ulem}
\usepackage[sort&compress,numbers]{natbib}
\hypersetup{colorlinks,citecolor= nicegreen,linkcolor= nicered}
\definecolor{nicered}{rgb}{0.7,0.1,0.1}
\definecolor{nicegreen}{rgb}{0.1,0.5,0.1}

\newcommand{\be}{\begin{equation}}
\newcommand{\ee}{\end{equation}}
\newcommand{\ba}{\begin{array}}
\newcommand{\ea}{\end{array}}
\newcommand{\bea}{\begin{eqnarray}}
\newcommand{\eea}{\end{eqnarray}}
\newcommand{\balg}{\begin{align}}
\newcommand{\ealg}{\end{align}}
\newcommand{\bit}{\begin{itemize}}
\newcommand{\eit}{\end{itemize}}
\newcommand{\trm}[1]{\textrm{#1}}

\newcommand{\Mpc}{\trm{\Mpc}}
\newcommand{\yr}{\trm{\yr}}
\newcommand{\eV}{\trm{\eV}}
\newcommand{\nn}{\nonumber}

\begin{document}

\hfill
NUHEP-TH/17-04

%\hfill \today

\vspace*{1.0cm}

\centerline{\bf\Large\textcolor[rgb]{0,0,0}{Dark Matter and Neutrino Mass}\vspace*{0.2cm}}
\centerline{\bf\Large\textcolor[rgb]{0,0,0}{from the Smallest Non-Abelian Chiral Dark Sector}}
\vspace*{1.2cm}

\centerline{\bf 
{Jeffrey M. Berryman\,\footnote{\tt jmberryman@u.northwestern.edu}}, \ \ 
{Andr\'e de Gouv\^ea\,\footnote{\tt degouvea@northwestern.edu}}, \ \  
{Kevin J. Kelly\,\footnote{\tt kjk@u.northwestern.edu}}, \ \  
{Yue Zhang\,\footnote{\tt yuezhang@northwestern.edu}}
}

\vspace*{1.0cm}

\centerline{\em Department of Physics \& Astronomy, Northwestern University}
\centerline{\em 2145 Sheridan Road, Evanston, IL 60208, USA}

\vspace{1.5cm}

\parbox{16.5cm}{
{\sc Abstract:} 
All pieces of concrete evidence for phenomena outside the standard model (SM) -- neutrino masses and dark matter -- are consistent with the existence of new degrees of freedom that interact very weakly, if at all, with those in the SM. We propose that these new degrees of freedom organize themselves into a simple dark sector, a chiral  $SU(3)\times SU(2)$ gauge theory with the smallest nontrivial fermion content. Similar to the SM, the dark $SU(2)$ is spontaneously broken while the dark $SU(3)$ confines at low energies.  At the renormalizable level, the dark sector contains massless fermions -- dark leptons -- and stable massive particles -- dark protons.  We find that dark protons with masses between 10--100~TeV satisfy all current cosmological and astrophysical observations concerning dark matter even if dark protons are a symmetric thermal relic. The dark leptons play the role of right-handed neutrinos and allow simple realizations of the seesaw mechanism or the possibility that neutrinos are Dirac fermions. In the latter case, neutrino masses are also parametrically different from charged-fermion masses and the lightest neutrino is predicted to be massless. Since the new ``neutrino'' and ``dark matter'' degrees of freedom interact with one another, these two new-physics phenomena are intertwined. Dark leptons play a nontrivial role in early universe cosmology while indirect searches for dark matter involve, decisively, dark matter annihilations into dark leptons. These, in turn, may lead to observable signatures at high-energy neutrino and gamma-ray observatories, especially once one accounts for the potential Sommerfeld enhancement of the annihilation cross-section, derived from the low-energy dark-sector effective theory, a possibility we explore quantitatively in some detail.
}

%%%%%%%%%%%%%%%%%%%%%%%%%%%%%%%%%%%%%%%%%%%%%%%%%%%%%%%%%%%%%%%%%%%%%%%%%%%%%%%%
\newpage

{\hypersetup{linkcolor=black}
\tableofcontents
}

%%%%%%%%%%%%%%%%%%%%%%%%%%%%%%%%%%%%%%%%%%%%%%%%%%%%%%%%%%%%%%%%%%%%%%%%%%%%%%%%

\setcounter{equation}{0}
\setcounter{footnote}{0}

%%%%%%%%%%%%%%%%%%%%%%%%%%%%%%%%%%%%%%%%%%%%%%%%%%%%%%%%%%%%%%%%%%%%%%%%%%%%%%%%

\newpage

\section{Introduction}

Nonzero neutrino masses imply the existence of fundamental particles beyond those that constitute the unreasonably successful standard model of particle physics (SM). The dark matter puzzle also strongly hints at the existence of new particles and new interactions. Both constitute the only unambiguous direct evidence that the SM is incomplete and are, unsurprisingly, the subject of intense theoretical and experimental investigation. Other than the fact that they exist, very little is known about these new degrees of freedom. They have never been directly observed in laboratories and are constrained to be very heavy or very weakly coupled. The ``parameter-spaces'' for the new physics responsible for nonzero neutrino masses and dark matter are immense. A priori, we don't know if the two problems are related, how the new degrees of freedom interact with the SM degrees of freedom, or how the new degrees of freedom interact with one another. 

Given the dearth of information, it is very tempting to extract inspiration from the SM. The SM is a chiral gauge theory, spontaneously broken via the Higgs mechanism to $SU(3)_c\times U(1)_{\rm EM}$ (strong interactions and electromagnetism). Given the SM gauge group, the particle content is, ignoring the fact there are three generations, minimal. If any of the known quarks or leptons failed the exist, the SM would be theoretically inconsistent and, for example, gauge invariance would be violated by quantum mechanical effects -- the gauge theory would be anomalous. There are interesting consequences to an anomaly-free, minimal, chiral gauge theory. All SM fermion masses are proportional to the electroweak symmetry breaking scale, and the SM Lagrangian contains accidental global symmetries that imply the existence of massive, stable matter particles (protons) and massless matter particles (neutrinos).\footnote{The electron is also a stable, massive particle. Its stability is guaranteed by electric charge conservation.} 

Here, we explore the possibility that the new degrees of freedom responsible for nonzero neutrino masses and dark matter are also described by a chiral gauge theory with spontaneous gauge symmetry breaking, for several reasons: (i) the particle content of chiral gauge theories is constrained but nontrivial, (ii) the new degrees of freedom interact with one another in a rigid and well prescribed way, (iii) all fundamental masses are governed by the scale of spontaneous symmetry breaking, (iv) the new degrees of freedom can be made to interact with the SM degrees of freedom in only a handful of different ways that are easy to parameterize -- the Higgs portal, the vector portal, and the neutrino portal -- and investigate phenomenologically, and (v) we expect, in general, new accidental global symmetries and with them new, massive stable particles and new massless fermions. These, it turns out, are excellent candidates for the dark matter particle and the degrees of freedom associated to nonzero neutrino masses. Collectively, we refer to the new degrees of freedom as the ``dark sector.''

It is not straightforward to construct chiral gauge theories because of anomaly cancellations. Several chiral $U(1)$ models have been investigated in the literature in the last several years~\cite{Babu:2004mj,Batra:2005rh,deGouvea:2015pea,Co:2016akw}, and many more have been identified~\cite{Batra:2005rh,deGouvea:2015pea}. A simple, general procedure for generating chiral $U(1)$ models was discussed in detail in Ref.~\cite{deGouvea:2015pea}. Abelian chiral dark sector models were discussed in Refs.~\cite{Babu:2004mj,deGouvea:2015pea,Co:2016akw}. In Ref.~\cite{Berryman:2016rot}, an equally simple, general procedure for generating non-abelian chiral gauge theories was presented, and the phenomenology of a few concrete models was discussed, briefly and qualitatively.  

More concretely, we explore the phenomenology of the smallest non-abelian chiral gauge theory that does not contain a $U(1)$ gauge group, which turns out to be an $SU(3)\times SU(2)$ gauge theory. We assume its symmetry breaking procedure to be similar to the SM: there is a dark Higgs doublet which breaks the dark $SU(2)$ and gives the fermions mass. The dark $SU(3)$ gauge coupling goes strong at a relatively lower energy scale where dark quarks confine into dark hadrons.\footnote{In the absence of a fundamental dark Higgs doublet, the $SU(2)$ would still be spontaneously broken together with chiral symmetry after the $SU(3)$ gauge theory confines. We will comment on this Higgsless case in Section~\ref{sec:conclusions}.} Details are discussed in Section~\ref{sec:SU3SU2}. On the other hand, the model is simpler than the SM. In particular, there are no gauged $U(1)$ symmetries. This is significant since,  

\begin{itemize}

\item The absence of a gauged $U(1)$ ``hypercharge'' implies there is no need of dark sector fields analogous to the SM right-handed charged leptons. As a result, the $SU(2)$ doublet dark leptons are massless at the renormalizable level and naturally serve as the partners (``right-handed neutrinos'') associated with nonzero neutrino masses. Details are discussed in Section~\ref{sec:neutrino}.

\item The absence of an unbroken $U(1)$ ``electromagnetic'' force -- a long range force -- implies it is possible to arrange the lightest dark baryon state (which is stable because of the accidentally conserved dark baryon number) to be a viable dark matter candidate even in the absence of a primordial dark baryon asymmetry. The dark baryons thermally freeze out as they annihilate into the lighter dark pions and the correct relic density, it turns out, points to dark matter masses at the 10--100 TeV scale. Details are discussed in Sections~\ref{sec:darkmatter} and \ref{sec:cosmo}. 

\item The absence of a gauged $U(1)$ implies the absence of the vector portal to the SM through kinetic mixing. Instead, we must resort to the Higgs or the neutrino portal in order to probe the dark sector. In particular, in Section~\ref{sec:detection}, we highlight the importance of the neutrino portal for dark matter indirect detection.

\end{itemize}

Throughout, we emphasize the tight connection between dark matter and the degrees of freedom related to nonzero neutrino masses. Both the dark quarks (constituents of the dark matter candidate) and the dark leptons (``right-handed neutrinos'' that couple to the active neutrinos) are required by the theoretical consistency of this smallest non-abelian chiral dark sector. In turn, these two species strongly influence one another's early universe dynamics. If the dark baryons are thermal relics, the entire dark sector, including the dark leptons, is necessarily in thermal equilibrium early enough in the history of the universe. This leads to strong constraints on the dark lepton parameters (masses and couplings) from cosmic surveys. Meanwhile, the dark leptons provide guidance regarding the parameters of the neutrino portal, impacting indirect searches for the dark matter as well as possible collider searches.

%%%%%%%%%%%%%%%%%%%%%%%%%%%%%%%%%%%%%%%%%%%%%%%%%%%%%%%%%%%%%%%%%%%%%%

\setcounter{footnote}{0}
\setcounter{equation}{0}
\section{The Model}
\label{sec:SU3SU2}

Following the results discussed in detail in Ref.~\cite{Berryman:2016rot}, $SU(3)\times SU(2)$ is the smallest, non-abelian, chiral gauge theory that does not contain a $U(1)$ gauge group. The minimal\footnote{Minimal refers to the smallest number of fermionic degrees of freedom necessary to render the theory anomaly free. Less minimal models would contain more chiral fields that transform under different representations or vector-like fermions.} fermion content is 
\begin{equation}
Q_D(3,2),~~~u^c_D(\bar{3},1),~~~d^c_D(\bar{3},1),~~~L_D(1,2),
\end{equation}
where all fermions are left-handed Weyl fermions and the symbols in parenthesis indicate how the different fields transform under $SU(3)\times SU(2)$ (for example, $Q_D$ transforms as a triplet under $SU(3)$ and a doublet under $SU(2)$). The dark quantum numbers of all the fields are identical to those of SM quarks and leptons under $SU(3)_c\times SU(2)_L$ (color and $SU(2)$ weak interactions) and hence we name the fields after their SM doppelg\"angers.  The subscript $D$ is present to eliminate confusion between dark sector fermions and those in the SM. 

We will consider only one generation of dark fermions. In this scenario, the chiral $L_D$ field is necessary in order to cancel the Witten anomaly \cite{Witten:1982fp} ($Q_D$ contains three $SU(2)$ doublets). The case of two generations is qualitatively different. If the number of generations is even, the number of $SU(2)$ doublets also charged under $SU(3)$ is even and the theory is anomaly free even in the absence of the $SU(3)$ singlets $L_D$. On the other hand, if there is more than one generation of dark fermions, the $L_D^i$ are allowed Dirac masses proportional to $\epsilon_{ab}L_D^aL_D^b$, where $a,b$ are generation indices, and hence have masses unrelated to $SU(2)$ symmetry breaking.    

It is easy to check that, unless the gauge symmetry is broken, all dark sector fundamental fermions are massless. We postulate that the $SU(3)\times SU(2)$ gauge symmetry is spontaneously broken via the Higgs mechanism. The breaking pattern and spectrum of fermion masses depend on the choice we make for the Higgs sector. For example, if the dark sector contains an anti-color-triplet scalar $T^c(\bar{3},1)$, the Lagrangian includes ${\cal L}\supset Q_DL_DT^c + D^c_DU^c_DT^c + Q_DQ_D(T^c)^{\dagger}+ {\rm h.c.}$, where the Yukawa couplings are omitted and the $SU(3)\times SU(2)$ contractions are implied. Were the $T^c$ scalar field to acquire a vacuum expectation value, part of the gauge symmetry would be spontaneous broken ($SU(3)\times SU(2) \to SU(2)\times SU(2)$) and a subset of the Weyl fermions would pair up into massive Dirac fermions, leaving behind a few massless Weyl fermions. Other choices include an $SU(2)$ scalar doublet $H_D(1,2)$, a bifundamental scalar $\Delta(3,2)$, and combinations of these three fields. 

Here, we postulate the existence of a single dark Higgs doublet in the scalar sector, $H_D(1,2)$, with a scalar potential that implies spontaneous symmetry breaking, $SU(3)\times SU(2) \to SU(3)$. The dark sector Lagrangian includes 
\begin{equation}
-{\cal L} \supset y_u Q_Du^c_DH_D + \tilde{y}_u Q_Du^c_D\tilde{H}_D+ y_dQ_Dd^c_DH_D + \tilde{y}_d Q_Dd^c_D\tilde{H}_D + {\rm h.c.} \ ,
\end{equation}
where $y_{u,d}$ and $\tilde{y}_{u,d}$ are Yukawa couplings and, as usual, $\tilde{H}_D\equiv i\sigma_2H_D^*$. Unlike the quarks in the SM, the up- and down-type dark quarks are allowed to mix after the symmetry breaking.

After spontaneous symmetry breaking
\begin{equation}
H_D \to \left(\begin{array}{c} 0 \\ {v_D}/{\sqrt2} \end{array}\right) \ ,
\end{equation}
and the three $SU(2)$ gauge bosons $X$ acquire identical masses, $M^2_X = g_2^2 v_D^2/4$, where $g_2$ is the $SU(2)$ gauge coupling. The mass matrix for the four dark quark Weyl fermions ($\times 3$ for dark color) is
\begin{equation}
{\cal M}_q =  
\left(\begin{array}{cc} y_u  & y_d  \\ \tilde{y}_u & \tilde{y}_d \end{array}\right) \frac{v_D}{\sqrt2}~.
\end{equation}
After diagonalization, they combine into two massive Dirac dark quarks $q_1$ and $q_2$. Unless otherwise noted, we will work on the mass eigenstate basis (masses $m_{q_1}$, $m_{q_2}$ with $m_{q_2}>m_{q_1}$) for the dark quarks. The dark quarks couple left-handedly to the $X$ gauge bosons. Because the three $X$ gauge boson masses are degenerate, we are free to redefine the generators of the broken dark $SU(2)$ and massive gauge boson fields such that, after spontaneous symmetry breaking, 
 \begin{equation}\label{LCC}
{\cal L} \supset \frac{g_2}{2\sqrt{2}}\left[\bar{q}_1\gamma_{\mu}(1-\gamma_5)q_2X^{\mu}_{+}+\bar{q}_2\gamma_{\mu}(1-\gamma_5)q_1X^{\mu}_{-}\right] + 
\frac{g_2}{4}\left[\bar{q}_1\gamma_{\mu}(1-\gamma_5)q_1-\bar{q}_2\gamma_{\mu}(1-\gamma_5)q_2\right]X_3^{\mu},
 \end{equation}
where $X^{\mu}_{\pm}=(X^{\mu}_1\mp iX^{\mu}_2)/\sqrt{2}$. In section~\ref{darkchpt}, we will introduce dark pions, composite states made of a dark quark and a dark antiquark. We will define $\{\pi^+_D, \pi^0_D, \pi^-_D\}=\left\{ q_1 \bar q_2, (q_1\bar q_1 - q_2 \bar q_2)/\sqrt2, q_2 \bar q_1 \right\}$, {\it i.e.}, the ``charge'' assignments correspond to the sign of the quark couplings to $X_3$ (third-component of dark $SU(2)$).

After spontaneous symmetry breaking, the two dark leptons in $L_D$, $\nu^c_1$ and $\nu^c_2$, remain massless. As we explore in Sec.~\ref{sec:neutrino}, these can play the role of left-handed antineutrinos. Because they are massless, one can define the dark lepton couplings to gauge bosons to agree with those of the dark quarks,
 \begin{equation}
{\cal L} \supset \frac{g_2}{2\sqrt{2}}\left[\bar{\nu}_1\gamma_{\mu}(1-\gamma_5)\nu_2X^{\mu}_{+}+\bar{\nu}_2\gamma_{\mu}(1-\gamma_5)\nu_1X^{\mu}_{-}\right] + 
\frac{g_2}{4}\left[\bar{\nu}_1\gamma_{\mu}(1-\gamma_5)\nu_1-\bar{\nu}_2\gamma_{\mu}(1-\gamma_5)\nu_2\right]X_3^{\mu}.
 \end{equation}
Here, for convenience, $\nu_1$ and $\nu_2$ are four-component Dirac fermions whose left-handed chiral components are $\nu^c_1$ and $\nu^c_2$ (and whose right-handed chiral components vanish).
 
After spontaneous symmetry breaking, the $SU(3)$ dark color gauge symmetry remains, along with a global dark baryon number $U(1)_{DB}$ and a global dark lepton number $U(1)_{DL}$. Parallel to SM quantum chromodynamics (QCD), the $SU(3)$ gauge symmetry confines and the dark quarks form dark baryons and dark mesons. The lightest dark baryon is guaranteed to be stable because of $U(1)_{DB}$ charge conservation and is an interesting dark matter candidate. We explore this possibility in detail in Secs. \ref{sec:darkmatter} and \ref{sec:cosmo}. 

The combined renormalizable Lagrangian describing the dark sector and the SM is
\begin{equation}
{\cal L} = {\cal L}_{SM}(Q,u^c,d^c,L,e^c,H) + {\cal L}_{DS}(Q_D,u^c_D,d^c_D,L_D,H_D) - \kappa|H|^2|H_D|^2.
\label{eq:Ltot}
\end{equation}
where $Q,u^c,d^c,L,e^c$ are the SM Weyl fermion fields (generation indices suppressed) and $H$ is the SM Higgs doublet. At the renormalizable level, the SM and the dark sector interact only via the Higgs portal, whose strength is governed by a single dimensionless coupling $\kappa$. In the limit $\kappa\to 0$, and ignoring gravity, the SM and the dark sector decouple.

%%%%%%%%%%%%%%%%%%%%%%%%%%%%%%%%%%%%%%%%%%%%%%%%%%%%%%%%%%%%%%%%%%%%%%

\setcounter{footnote}{0}
\setcounter{equation}{0}
\section{Dark Leptons -- The Neutrino Sector}
\label{sec:neutrino}

In the SM, neutrino masses are zero. The SM degrees of freedom and gauge symmetry do not allow, at the renormalizable level, neutrino masses even after symmetry breaking. The same happens in the dark sector; the dark sector degrees of freedom and gauge symmetry do not allow, at the renormalizable level, masses for $\nu_1^c$ and $\nu_2^c$, the left-handed antineutrinos.

The following field binomials are gauge invariant but not Lorentz invariant:\footnote{Many of the general features discussed here do not depend on the details of the dark symmetry breaking sector. Had we chosen a different dark Higgs sector, there would still be dark gauge-invariant binomials. For example, if the dark Higgs sector consisted of a dark color triplet $t(3,1)$, $u_D^ct$ and  $d_D^ct$ would be $SU(3)\times SU(2)$ gauge invariants. In this case, some of the components of $u^c_D$ and $d^c_D$, massless in the absence of higher dimensional operators, would play the role of the left-handed antineutrinos.} $LH$, $L_DH_D$, $L_D\tilde{H}_D$. Pairs of those can be chosen as Lorentz and gauge invariant, and lead to the following dimension-five Lagrangian. Note that this is the most general dimension-five Lagrangian consistent with SM and dark gauge invariance. 
\begin{eqnarray}
 \label{eq:dim5}
{\cal L}_5 & = & -\frac{y_{11}}{2\Lambda_{\nu}}(L_DH_D)(L_DH_D) -\frac{y_{22}}{2\Lambda_{\nu}}(L_D\tilde{H}_D)(L_D\tilde{H}_D) -\frac{y_{12}}{\Lambda_{\nu}}(L_DH_D)(L_D\tilde{H}_D) + H.c. \nonumber \\
& & -\frac{y_{\alpha\beta}}{2\Lambda_{\nu}}(L_{\alpha}H)(L_{\beta}H) + H.c.  \\
& & -\frac{y_{1\alpha}}{\Lambda_{\nu}}(L_{\alpha}H)(L_DH_D) -\frac{y_{2\alpha}}{\Lambda_{\nu}}(L_{\alpha}H)(L_D\tilde{H}_D) + H.c., \nonumber
\end{eqnarray}
where $y$'s are dimensionless couplings ($y_{\alpha\beta}=y_{\beta\alpha}$), $\Lambda_{\nu}$ is the effective scale associated with the effective operators, and $\alpha, \beta=e,\mu,\tau$ runs over the three SM lepton flavors. Without loss of generality, we associate the same effective scale $\Lambda_{\nu}$ to all dimension-five operators and allow the $y$ couplings to be hierarchical in order to take into account that potentially different new physics effects may be responsible for the different operators. The list of operators in Eq.~(\ref{eq:dim5}) could be obtained from integrating out heavy new gauge-singlet fermions. If, on the contrary, the gauge-singlet fermions are light, the neutrino sector could be more complicated as, for example, in the inverse-seesaw scenario~\cite{Mohapatra:1986aw}.

The first line in Eq.~(\ref{eq:dim5}) explicitly violates the dark lepton-number symmetry. After spontaneous symmetry breaking, it endows $\nu_1^c$ and $\nu_2^c$ with Majorana masses proportional to $v_D^2/\Lambda_{\nu}$. The left-handed antineutrino Majorana mass matrix is
\begin{equation}
M_{RR} = \left(\begin{array}{cc} y_{11} & y_{12} \\ y_{12} & y_{22 }\end{array}\right)\frac{v_D^2}{2\Lambda_{\nu}}.
\end{equation} 

The second line in Eq.~(\ref{eq:dim5}) is the well known Weinberg operator and explicitly violates SM lepton number $U(1)_{\ell}$. After spontaneous symmetry breaking, it endows the SM neutrinos with Majorana masses proportional to $v^2/\Lambda_{\nu}$, where $v/\sqrt2$ is the expectation value of the neutral component of the SM Higgs field, $v=246\,$GeV. The elements of the SM neutrino Majorana mass matrix are, in the flavor basis, $(M_{LL})_{\alpha\beta}=y_{\alpha\beta}v^2/2\Lambda_{\nu}$.

The third line in Eq.~(\ref{eq:dim5}) explicitly violates the global lepton number symmetries in both the SM and the dark sector. It does, however, preserve a diagonal subgroup $U(1)_{L}$ under which SM leptons and the left-handed antineutrino fields have equal and opposite charges. After spontaneous symmetry breaking, it leads to Dirac masses to the two $\nu^c$ fields and two linear combinations of SM neutrinos.  The $3\times 2$ Dirac neutrino mass matrix is
\begin{equation}
M_{RL} = \left(\begin{array}{cc} y_{1e} & y_{2e}  \\ y_{1\mu} &  y_{2\mu} \\  y_{1\tau} & y_{2\tau} \end{array}\right)\frac{vv_D}{2\Lambda_{\nu}}.
\end{equation} 

In summary, the situation is as follows. At the renormalizable level, SM neutrinos and the dark sector left-handed antineutrinos are massless. More new physics (we refer to it as ``$\nu$physics'' in order to distinguish it from the dark sector) is required in order to render the the SM neutrinos massive. If the effects of the new physics can be captured by higher dimensional operators associated to a $\nu$physics effective scale $\Lambda_{\nu}$, several options emerge, depending on whether the $\nu$physics preserves different accidental global symmetries of the Lagrangian. 

If the $\nu$physics does not preserve any of the accidental lepton-number symmetries of the SM plus dark sector, the $5\times 5$ Majorana neutrino mass matrix is, after spontaneous symmetry breaking,
\begin{equation}
M_{\nu} = \left(\begin{array}{cc} M_{LL} & M_{RL} \\ M_{RL}^T & M_{RR} \end{array}\right).
\label{eq:seesaw}
\end{equation}
In general, there are five massive Majorana neutrinos, all of them linear combinations of the three SM flavors plus $\nu_1^c$ and $\nu_2^c$. Excluding the possibility that the neutrinos are pseudo-Dirac fermions, which we do not explore here, three of the neutrino mass eigenstates must be predominantly composed of $\nu_e, \nu_{\mu}, \nu_{\tau}$ in accordance with the world's neutrino data. These mass eigenstates will be referred to as the mostly-active neutrinos. The other two mass eigenstates are predominantly linear combinations of the dark left-handed antineutrinos $\nu_1^c$ and $\nu_2^c$. Hereafter, these mass eigenstates will be referred to as the mostly-sterile neutrinos $(\nu_D)_{1,2}$ or the  ``dark leptons.''\footnote{We use both designations -- mostly-sterile neutrinos and dark leptons -- interchangeably.} The associated phenomenology depends on the relative magnitudes of $M_{LL}$, $M_{RR}$, and $M_{RL}$ and some of it has been extensively discussed in the literature in, for example, mirror-world scenarios \cite{Akhmedov:1992hh,Berezhiani:1995yi,An:2009vq}. The neutrino sector mass spectrum and mixing can be classified into the following cases:

{\it Heavy mostly-sterile neutrinos.} In this case, the mass matrices satisfy the hierarchy $M_{LL}, M_{RL}\ll M_{RR}$. The mostly-active neutrinos receive their masses from two sources, $M_{LL}$, and a type-I seesaw contribution, $-(M_{RL})^2/M_{RR}$, obtained from integrating out the left-handed antineutrinos. The mostly-sterile neutrinos, with masses proportional to $M_{RR}$,  manifest themselves as new physics particles often denominated neutral heavy leptons. The active--sterile mixing angle $\theta_{as}$ depends on the relative size of $M_{LL}$ and $M_{RL}^2/M_{RR}$. If $M_{LL}\gg M_{RL}^2/M_{RR}$, $\theta_{as}\sim M_{RL}/M_{RR} < \sqrt{m_\nu/m_{\nu_D}}$. Under these conditions, the dark sector is not directly responsible for generating the observed nonzero neutrino masses.

On the other hand, if $M_{LL}\ll M_{RL}^2/M_{RR}$, Eq.~(\ref{eq:seesaw}) is well approximated by the renowned Type-I seesaw mechanism (with two right-handed neutrinos). Generic predictions include, in the limit $M_{LL}\to 0$, one massless active neutrino, two mostly-active massive states with masses of order $(M_{RL})^2/M_{RR}$, and two mostly-sterile states with masses of order $M_{RR}$. Active--sterile mixing is of order $\theta_{as}\sim M_{RL}/M_{RR} \simeq \sqrt{m_\nu/m_{\nu_D}}$. 

Another intriguing possibility is $v/v_D\ll 1$ while all dimensionless couplings $y$ in Eq.~(\ref{eq:dim5}) are of the same magnitude. In this case, there are two heavy mostly-sterile states (with masses of order $M_{RR}$), but $M_{LL}$ is of order the Type-I seesaw contribution to the mostly-active neutrino masses, $(M_{RL})^2/M_{RR}$. In this case, the relationship between the neutrino masses and the active--sterile mixing parameters is blurred and there is, for example, the possibility of having $\theta_{as}\sim M_{RL}/M_{RR} > \sqrt{m_\nu/m_{\nu_D}}$.

{\it Light (eV-scale) mostly-sterile neutrinos.} If all $M_{LL}$, $M_{RR}$, and $M_{RL}$ elements are of the same magnitude (this happens if all dimensionless couplings are similar and $v/v_D\sim 1$), we expect two very light mostly-sterile neutrinos that mix significantly with the mostly-active neutrinos. This scenario is actively being explored, for example, in oscillation searches for new neutrino mass eigenstates and could provide a solution to the short-baseline anomalies~\cite{deGouvea:2005er}.\footnote{There may be severe constraints from cosmology if the mostly-sterile neutrinos have eV scale masses and sizable active--sterile mixing angles, in which case they can be over-produced via neutrino oscillations in the early universe~\cite{Hamann:2011ge,Hernandez:2013lza}. In Sec.~\ref{sec:cosmo}, we explore cosmological implications of eV-scale mostly-sterile neutrinos. There we implicitly assume the active--sterile mixing angles small enough that the production of mostly-sterile neutrinos via oscillations is negligible.}

{\it  Decoupled sterile neutrinos.} If the $\nu$physics preserves the dark sector lepton-number, $M_{RL}$ and $M_{RR}$ vanish.\footnote{If the $\nu$physics also preserves the SM $U(1)_{\ell}$, SM neutrinos remain massless. We ignore this possibility.} In this case, the left-handed antineutrinos are massless and completely decoupled from the neutrinos or all other SM degrees of freedom, $\theta_{as}=0$. They may still, however, play a significant cosmological role as relics of the Big Bang, as we discuss in the next section.

{\it  Dirac neutrinos.} If the $\nu$physics violates both the dark sector and the SM lepton numbers but preserves the diagonal $U(1)_L$ subgroup, both $M_{LL}$ and $M_{RR}$ vanish. In this case, neutrinos are Dirac fermions and the Dirac mass matrix is given by $M_{RL}$; furthermore, one of the neutrino masses is zero. It is amusing that, if the neutrinos are Dirac fermions, only one dark-sector-family already contains two right-handed neutrino degrees of freedom and is sufficient to accommodate our understanding of neutrino masses and lepton mixing. In this case, the connection between the dark sector and nonzero neutrino masses is strongest; the dark sector provides all the degrees of freedom required to allow (two of) the neutrinos to be massive Dirac fermions. Even though neutrinos are Dirac fermions, their masses are parametrically different from those of all other SM and dark sector fields. The neutrino Dirac masses are proportional to $vv_D/\Lambda_{\nu}$, while the masses of charged SM (other dark sector) objects are proportional to $v$ ($v_D$). In the limit $\Lambda_{\nu}\gg v,v_D$, one expects neutrino masses to be much smaller than those of all other fermions. Numerically, 
\begin{equation}
(m_{\nu})_{\rm Dirac} \sim 0.1~{\rm eV}\left(\frac{v_D}{10^3~\rm TeV}\right)\left(\frac{10^{15}~\rm TeV}{\Lambda_{\nu}}\right) \ . 
\end{equation}
If the neutrinos are Dirac fermions, the observed neutrino mass scale can be obtained if the effective $\nu$physics scale $\Lambda_{\nu}$ is around the Planck scale and $v_D$ is at the PeV scale, of interest to our dark matter considerations (see Sec.~\ref{sec:cosmo}).  

Before proceeding, since nonzero neutrino masses require the existence of degrees of freedom beyond those of the SM and the dark sector, it is interesting to explore, briefly and generically, other potential effects of the $\nu$physics, at least as far as those can be captured by higher dimensional operators. If $\Lambda_{\nu}$ is very high, most other higher dimensional operators, of dimension six and higher, will not lead to any observable effects, but there are exceptions. In the SM, for example, dimension six baryon number plus lepton number violating operators ($QQQL$, etc) lead to nucleon decay and are severely constrained. Equivalently, there are such operators in the dark sector ($Q_DQ_DQ_DL_D$) and these mediate the decay of dark baryons into dark mesons and dark leptons. It is also possible to write down ``mixed'' baryon-plus-lepton-number-violating operators involving particles from both sectors, such as $(u^c u^c d^c)(L_D H_D)$. These may play a role in generating the baryon asymmetry in the universe~\cite{Davoudiasl:2015jja,Dev:2015uca}. Whether the $\nu$physics has anything to do with baryon number violation (of the SM or the dark sector kind) is model dependent and beyond the aspirations of this manuscript. Henceforth, we ignore the possibility of all such effects.

\subsection{Decay of heavy dark leptons (mostly-sterile neutrinos)}

If the neutrinos are Majorana fermions, the mostly-sterile $\nu_D$ mass eigenstates (dark leptons) are allowed to be much heavier than the mostly-active ones (the {\it ``Heavy mostly-sterile neutrinos''} case defined above). Furthermore, if there is active--sterile mixing (i.e., if $M_{RL}\neq 0$), these states are necessarily unstable. Finally, assuming they are the lightest dark sector fermions, they can only decay into SM degrees of freedom.\footnote{We neglect the possibility of the heavier dark lepton decaying into three copies of the lighter dark lepton, a process that would occur, if kinematically allowed, even in the absence of active--sterile mixing. If this decay channel is relevant, the discussion in this subsection still applies to the lighter of the two dark leptons.} $\nu_D$ lifetimes are governed by their masses and the ``active--sterile'' elements of the neutrino mixing matrix, $U_{\alpha 4}$ and $U_{\alpha 5}$ (of order the mixing parameter $\theta_{as}$ define above), with $\alpha = e,\mu,\tau$.\footnote{We order the neutrino mass eigenstates so $\nu_4$ and $\nu_5$ are $(\nu_D)_{1,2}$, assumed to be much heavier than the other three.} In the next section, we will be interested in the fate of the relic dark leptons and hence need to understand quantitatively how long-lived they can be.

\begin{figure}[t]
\centerline{\includegraphics[scale=0.52]{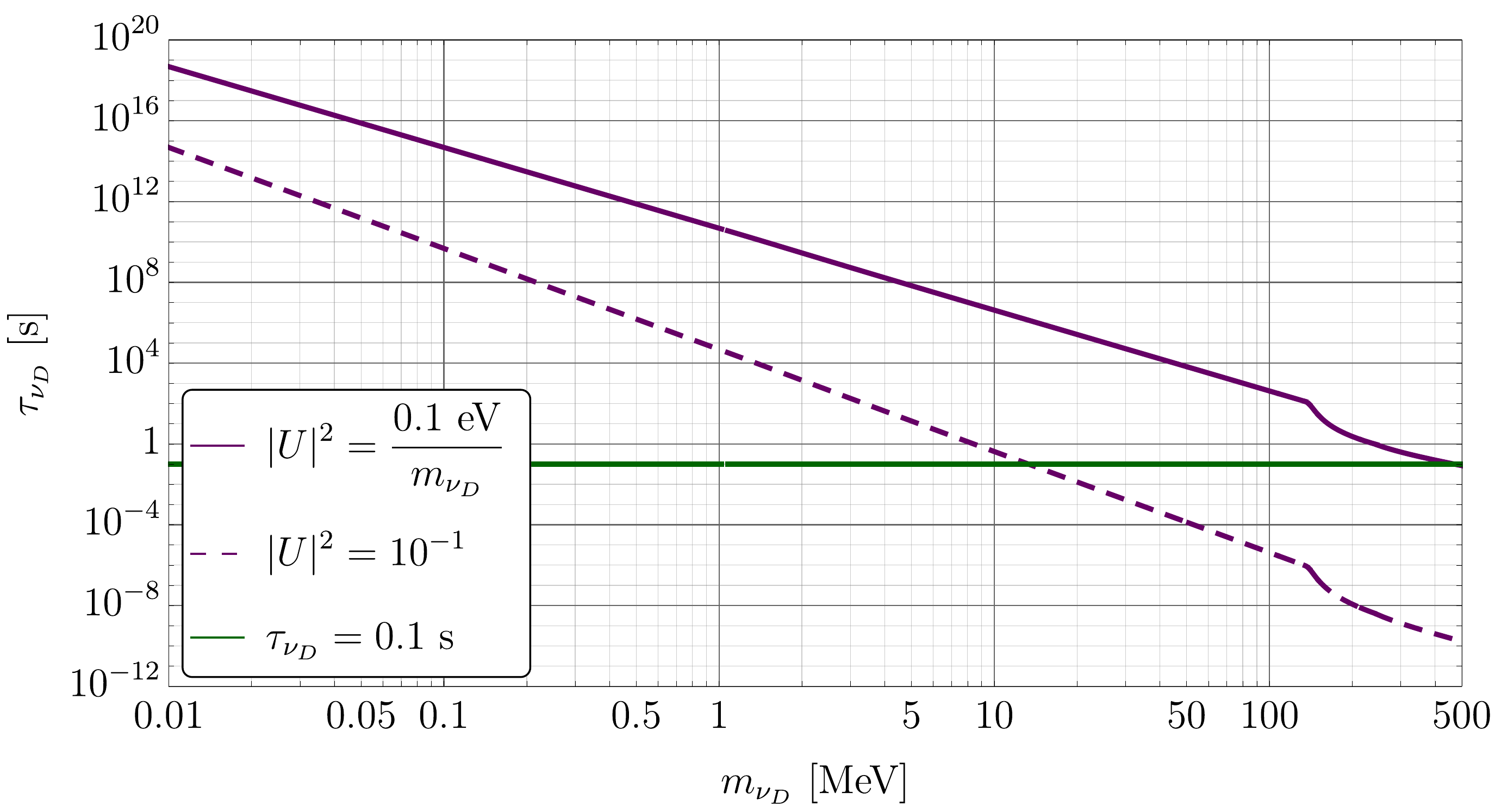}}
\caption{The mostly-sterile neutrino lifetime, $\tau_{\nu_D}$, as a function of its mass, $m_{\nu_D}$. The solid purple curve corresponds to $|U|^2 = {0.1 \text{ eV}}/{m_{\nu_D}}$, in agreement with general expectations from the Type-I seesaw scenario. The dashed purple curve corresponds to $|U|^2 = 10^{-1}$, representative of the weakest upper bound on active--sterile mixing in the mass range of interest. The solid green line shows $\tau_{\nu_D} = 0.1$ s, which is the upper bound on mostly-sterile neutrino decays at the time of BBN calculated in Ref.~\cite{Dolgov:2000jw}. For more details see Appendix~\ref{sec:SterileNuAppendix}.}\label{fig:nulife}
\end{figure}

Very light dark leptons (with mass below 1\,MeV) decay into three mostly-active neutrinos or into a mostly-active neutrino plus a photon. The associated decay widths are a steep function of the mass (proportional to $m^5$) and the magnitude-squared of the active--sterile neutrino mixing parameters, $|U_{\alpha4}|^2,|U_{\alpha5}|^2$. Heavier dark leptons can decay into charged-leptons and mostly-active neutrinos (similar to the case of muon decay, these are tree-body final states and the decay rate is proportional to $m^5$). For dark lepton masses above the pion mass, they can also decay into pions and charged-leptons or mostly-active neutrinos (similar to tau decay, the rates for these two-body decays are proportional to $m^3$). We provide more information on the branching ratios of the dark lepton decays in Appendix~\ref{sec:SterileNuAppendix}.

Fig.~\ref{fig:nulife} depicts the lifetimes of the mostly-sterile neutrinos as a function of their mass $m_{\nu_D}$, for different assumptions regarding active--sterile mixing. The neutrinos are considered to be Majorana fermions and we assume $|U_{\alpha 4}|^2=|U_{\alpha 5}|^2\equiv |U|^2$ for all $\alpha=e,\mu,\tau$. Bounds on active--sterile neutrino mixing as a function of the dark lepton masses have been compiled recently by several authors; see, for example, Refs.~\cite{Smirnov:2006bu,Atre:2009rg,Drewes:2015iva,Deppisch:2015qwa,deGouvea:2015euy,Alekhin:2015byh,Fernandez-Martinez:2016lgt,Adhikari:2016bei,Abada:2016plb} and references therein. In the bottom curve, $|U|^2$ is set to $|U|^2=10^{-1}$, representative of the weakest upper bounds on active--sterile mixing in the mass range of interest. This curve, therefore, represents an absolute lower bound on the dark lepton lifetimes. In the top curve, $|U|^2$ is set to $0.1~{\rm eV}/m_{\nu_D}$, in agreement with general expectations of the standard Type-I seesaw. Generically, we expect lifetimes between these two curves, while longer lifetimes remain a possibility in some scenarios. Fig.~\ref{fig:nulife} reveals that, depending on their masses, the dark lepton lifetimes vary from a fraction of a second to thousands of years to longer than the age of the Universe. With masses above 500~MeV, dark leptons are expected to decay in under a tenth of a second and those produced in the early universe will safely decay away before SM neutrinos decouple from the SM thermal bath. Dark leptons with masses above 100~keV are likely to decay before the formation of the cosmic microwave background (CMB). Mostly-sterile neutrinos lighter than tens of eV are expected to live longer than the age of the universe.  

Mostly-sterile neutrino lifetimes and the associated phenomenologies were explored extensively in the literature~\cite{Shrock:1980ct,Shrock:1981wq,Pal:1981rm,Johnson:1997cj,Gorbunov:2007ak,Ballett:2016opr}, and the results presented here are based mostly on Refs.~\cite{Pal:1981rm,Gorbunov:2007ak}. The green horizontal line in Fig.~\ref{fig:nulife} corresponds to $\tau_{\nu_D}=0.1$~s, consistent with the upper bound on mostly-sterile neutrino decays from big-bang nucleosynthesis (BBN) computed in Ref.~\cite{Dolgov:2000jw}.\footnote{Keep in mind that these bounds do not translate directly into the scenario under consideration here.} We comment on cosmological constraints on the dark leptons in Sec.~\ref{sec:cosmo}. 

%%%%%%%%%%%%%%%%%%%%%%%%%%%%%%%%%%%%%%%%%%%%%%%%%%%%%%%%%%%%%%%%%%%%%%

\setcounter{footnote}{0}
\setcounter{equation}{0}
\section{Dark Baryons and Mesons -- The Dark Matter Sector}
\label{sec:darkmatter}

After spontaneous $SU(2)$ symmetry breaking in the dark sector, the fermions charged under dark $SU(3)$ -- dark quarks -- acquire nonzero masses and are described as two massive dark-color triplet Dirac fermions $q_1$ and $q_2$, with masses $m_{q_1}$ and $m_{q_2}$, as described in Sec.~\ref{sec:SU3SU2}. The particle content is such that the unbroken $SU(3)$ gauge symmetry is ultraviolet free and hence confines at low energies (indeed, the particle content is identical to $SU(3)_c$ in the SM, if there was only one generation of SM fermions). At low energies, the propagating degrees of freedom are dark-color neutral bound states of dark quarks -- dark baryons ($q_iq_jq_k$), dark mesons ($\bar{q}_iq_j$), dark glueballs, etc. ($i,j,k=1,2$)\footnote{For recent work on composite dark sectors with a different model, see Ref.~\cite{Antipin:2015xia}.}.  

Two unrelated mass scales define the physics of the dark hadrons: the confinement scale $\Lambda_3$ and the dark quark masses (unless otherwise noted, we assume $m_{q_1}\sim m_{q_2}$). The dark quark masses, in turn, are related to $v_D$, the $SU(2)$ symmetry breaking scale, which also governs some of the properties of the dark hadrons, as we discuss in detail below. The phenomenology depends on the relative ordering of $v_D$, $\Lambda_3$ and $m_{q_{1,2}}$. Here, we restrict the discussion to scenarios where $v_D\gg \Lambda_3 \gg m_{q_{1,2}}$, for two reasons. One is we find that this choice provides a solution to the dark matter puzzle, and the other is these scenarios resemble the strong interactions in the SM (with only the first generation of quarks) so we can translate some of our understanding and intuition of low energy hadronic physics to the dark sector.

Assuming $v_D\gg \Lambda_3 \gg m_{q_{1,2}}$, we can readily identify the following. Similar to the SM, chiral symmetry is softly broken by the small but nonzero dark quark masses so there are three pseudo-goldstone bosons -- dark pions $\pi_D$ -- with masses of order $m_{\pi_D}\sim \sqrt{m_{q_{1,2}}\Lambda_3}$. The dark pions decay, exclusively, ``weakly'' into dark leptons, as will be discussed in detail in Sec.~\ref{sec:cosmo}. A fourth failed pseudo-goldstone boson, the $\eta'_D$, has mass of order $\Lambda_3$ and decays into dark pions. All other mesons, including dark glueballs, have masses of order $\Lambda_3$ and also decay promptly into dark pions.  

The lightest dark baryon is stable due to the accidental global $U(1)_{DB}$ symmetry. We refer to this state as the dark proton, $p_D$. Other dark baryon states will decay, if kinematically allowed, into the dark proton plus dark pions. States that are close in mass to the dark proton, if any, will decay ``weakly'' into the dark proton plus dark leptons.\footnote{Unless otherwise noted, we assume the dark leptons are much lighter than the dark hadrons independent of whether they are Majorana or Dirac fermions, see Sec.~\ref{sec:neutrino}.} For example, the second-lightest dark baryon, the dark neutron $n_D$, may only be allowed to decay into the dark proton and the dark leptons: $n_D \to p_D\nu_D\bar{\nu}_D$. Here, for simplicity, we assume the dark quark masses are degenerate enough that the lightest dark baryon states, the dark proton and neutron, have spin one-half.

\subsection{Interactions between dark baryons and mesons}\label{darkchpt}

The small dark quark masses imply an approximate $SU(2)_L\times SU(2)_R$ (global) symmetry, which is spontaneously broken by the dark quark condensate. The low energy effective theory describing the pseudo-goldsone bosons and the baryons in the dark sector is analogous to chiral perturbation theory in SM QCD. The leading nucleon-pseudoscalar couplings can be described following Ref.~\cite{Jenkins:1991ts},
\begin{eqnarray}
\mathcal{L}_{\rm D\chi PT} \supset \bar N i \gamma^\mu (\partial_\mu +v_\mu) N + g_A \bar N \gamma^\mu \gamma_5 a_\mu N \ ,
\end{eqnarray}
where
\begin{eqnarray}
\begin{split}
&v_\mu = \frac{1}{2} \left( u \partial_\mu u^\dagger + u^\dagger \partial_\mu u  \right), \ \ \ 
a_\mu = \frac{i}{2} \left( u \partial_\mu u^\dagger - u^\dagger \partial_\mu u  \right) \ , \\
&u = \sqrt{U} = e^{i \Pi/(2F_{\pi_D})}, \ \ \ \Pi =
\begin{pmatrix}
\pi_D^0 + \eta_D' & \sqrt{2} \pi_D^+ \\
\sqrt{2}\pi_D^- & -\pi_D^0 + \eta_D'
\end{pmatrix}, \ \ \ 
N = \begin{pmatrix}
p_D \\
n_D 
\end{pmatrix}~.
\end{split}
\end{eqnarray}
Here, keeping in mind there is no dark-sector equivalent of SM electromagnetism, the $\pi_D^{\pm, 0}$ are defined as prescribed in Eq.~(\ref{LCC}). In terms of the dark quark mass eigenstates, $\{\pi^+_D, \pi^0_D, \pi^-_D\} = \left\{ q_1 \bar q_2, (q_1\bar q_1 - q_2 \bar q_2)/\sqrt2, q_2 \bar q_1 \right\}$.

The interactions involving one or two pseudoscalar fields are
\begin{eqnarray}\label{Lchpt}
\begin{split}
\mathcal{L}_{\rm D\chi PT} &\supset \frac{g_A}{2F_{\pi_D}} \left[ \bar p_D \gamma^\mu \gamma_5 p_D (\partial_\mu \pi_D^0 + \partial_\mu \eta_D') + 
\bar n_D \gamma^\mu \gamma_5 n_D ( - \partial_\mu \pi_D^0 + \partial_\mu \eta_D') \rule{0mm}{4mm}\right. \\
& \hspace{1.6cm} + \left. \sqrt{2} \bar n_D \gamma^\mu \gamma_5 p_D \partial_\mu \pi_D^- 
+ \sqrt{2} \bar p_D \gamma^\mu \gamma_5 n_D \partial_\mu \pi_D^+ \rule{0mm}{4mm}\right] \\
&+ \frac{g_V^2}{F_{\pi_D}^2} \left[ \frac{i}{4} \bar p_D \gamma^\mu p_D ( \pi_D^+ \partial_\mu \pi_D^- - \pi_D^- \partial_\mu \pi_D^+) - \frac{i}{4} \bar n_D \gamma^\mu n_D ( \pi_D^+ \partial_\mu \pi_D^- - \pi_D^- \partial_\mu \pi_D^+)  \right. \\
& \left. \hspace{1.6cm} + \frac{i}{2\sqrt{2}} \bar n_D \gamma^\mu p_D ( \pi_D^- \partial_\mu \pi_D^0 - \pi_D^0 \partial_\mu \pi_D^-) + \frac{i}{2\sqrt{2}} \bar p_D \gamma^\mu n_D ( - \pi_D^+ \partial_\mu \pi_D^0 + \pi_D^0 \partial_\mu \pi_D^+) \right] \ ,
\end{split}
\end{eqnarray}
where we have included the parameter $g_V\equiv1$ so the following results appear more symmetric. These interactions are relevant for calculating the dark nucleon annihilation cross-section and thermal freeze-out in Section~\ref{TFO}.

The Lagrangian above ignores contributions from the CP-violating dark $\theta_D$-term
\begin{eqnarray}\label{Theta-Term}
\frac{\theta_D}{32\pi^2} (G_D)^a_{\mu\nu} (\tilde{G}_D)^{a \mu\nu} \ ,
\end{eqnarray} 
where $G_D$ contains the dark $SU(3)$ gauge bosons. We discuss other consequences of $\theta_D$ in the next subsection.  Here, we assume $\theta_D$ small enough that the trilinear $\eta'_D\pi_D \pi_D$ coupling is suppressed.  $\theta_D$-corrections to the annihilation cross-sections computed in the next section  (see Eq.~(\ref{Xsecs})) are of order $\theta_D^2$ and will be neglected. Note that the parity-violating Wess-Zumino-Witten term also does not contain couplings of three pseudoscalar particles~\cite{Scherer:2002tk}.

\subsection{$\theta_D$-term induced mixing between the dark pion and dark Higgs}

The most important effect of the $\theta_D$ term for dark sector phenomenology is to induce a mixing between the dark pion and the dark Higgs boson. The leading chiral Lagrangian for $\pi_D$ and $\eta'_D$ including the $\theta_D$-term takes the form
\begin{eqnarray}
\mathcal{L}_{\rm D\chi PT} \supset \frac{F_{\pi_D}^2}{4} {\rm Tr} \left( \partial_\mu U^\dagger \partial^\mu U \right) + B {\rm Tr} \left( M_q U + U^\dagger M_q^\dagger \right) - C \left( -i \log \det U - \theta_D \right)^2 \ ,
\end{eqnarray}
where $M_q = {\rm diag}\{m_{q_1}, m_{q_2}\}$ are the dark quark masses and the parameters $B, C>0$ control the dark pion and eta-prime masses (see Eq.~(\ref{BC})). Focusing on the  ``neutral'' fields, the $U$ matrix can be written as
\begin{eqnarray}
U = \left(\cos\frac{\pi_D^0}{F_{\pi_D}} + i \sin \frac{\pi_D^0}{F_{\pi_D}} \sigma_3 \right) \left(\cos\frac{\eta'_D}{F_{\pi_D}} + i \sin \frac{\eta'_D}{F_{\pi_D}} \right) \ ,
\end{eqnarray}
and the potential terms involving $\pi^0, \eta'$ are
\begin{align}\label{potential}
V(\pi^0_D, \eta'_D) =  & - B \left[ (m_{q_1} + m_{q_2}) \cos\frac{\pi_D^0}{F_{\pi_D}} \cos\frac{\eta'_D}{F_{\pi_D}} - (m_{q_1} - m_{q_2}) \sin\frac{\pi_D^0}{F_{\pi_D}} \sin\frac{\eta'_D}{F_{\pi_D}} \right] \nonumber \\
& + \frac{4C}{F_{\pi_D}^2} \left( \eta'_D - \frac{F_{\pi_D}\theta_D}{2} \right)^2 .
\end{align}
To minimize the potential, it is important to notice that the $\pi^0_D$ gets its mass from the $B$-term, while the $\eta'_D$ meson mainly gets its mass from the $C$-term. As a result, the vacuum expectation value (vev) of the $\eta'_D$ is dictated by the $C$-term at leading order in an $m_{\pi_D}^2/m_{\eta'_D}^2$ expansion, {\it i.e.}, $\langle \eta'_D \rangle/F_{\pi_D} =\theta_D/2$. 

After fixing the vev of the $\eta'_D$, the potential can be written as
\begin{eqnarray}
V(\pi^0_D, \eta'_D) = -B \left[ (m_{q_1} + m_{q_2}) \cos\frac{\pi_D^0}{F_{\pi_D}} \cos\frac{\theta_D}{2} - (m_{q_1} - m_{q_2}) \sin\frac{\pi_D^0}{F_{\pi_D}} \sin\frac{\theta_D}{2} \right]  \ .
\end{eqnarray}
Minimizing it with respect to the $\pi_D^0$ field, we obtain
\begin{eqnarray}
\tan \frac{\langle \pi_D^0 \rangle}{F_{\pi_D}}  = \frac{m_{q_2} - m_{q_1}}{m_{q_1} + m_{q_2}} \tan \frac{\theta_D}{2} \ .
\end{eqnarray}
This result agrees with Eq.~(20) of Ref.~\cite{Witten:1980sp}, where $\phi_{u,d}\equiv \mp{\langle \pi^0 \rangle}/{F_{\pi_D}} + \theta_D/2$.

Next, we shift the fields in order to express them as excitations about their vev's,  $\pi^0_D \to \langle \pi^0_D \rangle + \pi^0_D$, $\eta'_D \to \langle \eta'_D \rangle + \eta'_D$. This yields the following quadratic terms in the potential:
\begin{eqnarray}
V \supset \frac{1}{2} \begin{pmatrix} \pi_D^0 & \eta_D' & h_D \end{pmatrix}
\begin{pmatrix} 
m_{\pi_D}^2 & \delta_{\pi_D \eta'_D}^2 & 0 \\
\delta_{\pi_D \eta'_D}^2 & m_{\eta'_D}^2 & \delta_{\eta'_D h_D}^2 \\
0 & \delta_{\eta'_D h_D}^2 & m_{h_D}^2
\end{pmatrix}
\begin{pmatrix} \pi_D^0\\ \eta_D'\\ h_D \end{pmatrix} \ ,
\end{eqnarray}
where we have expanded the quark masses in terms of the dark Higgs field, $m_{q_{1,2}}\to m_{q_{1,2}}(1+ h_D/v_D)$, and
\begin{eqnarray}\label{BC}
\begin{split}
&m_{\pi_D}^2 = \frac{2B}{F_{\pi_D}^2} \sqrt{m_{q_1}^2 + m_{q_2}^2 + 2 m_{q_1} m_{q_2} \cos\theta_D} \ , \\
&m_{\eta'_D}^2 = - \frac{8C}{F_{\pi_D}^2} +  \frac{2B}{F_{\pi_D}^2} \sqrt{m_{q_1}^2 + m_{q_2}^2 + 2 m_{q_1} m_{q_2} \cos\theta_D} \ , \\
&\delta_{\pi_D\eta'_D}^2 = - \frac{2B}{F_{\pi_D}^2} \frac{m_{q_2}^2 - m_{q_1}^2}{\sqrt{m_{q_1}^2 + m_{q_2}^2 + 2 m_{q_1} m_{q_2} \cos\theta_D}} \ , \\
&\delta_{\eta'_Dh_D}^2 = \frac{8B}{F_{\pi_D} v_D} \frac{m_{q_2} m_{q_1} \tan ({\theta_D}/{2})}{\sqrt{m_{q_1}^2 + m_{q_2}^2 + 2 m_{q_1} m_{q_2} \cos\theta_D}} \ .
\end{split}
\end{eqnarray}

In the limit $\delta_{\pi_D\eta'_D}^2 \ll m_{\eta'_D}^2$ and $\delta_{\eta'_Dh_D}^2 \ll m_{h_D}^2$, and $\theta_D\ll1$, we find
\begin{eqnarray}\label{darkpionHiggs}
\theta_{\pi^0_D h_D} \simeq \theta_{\pi^0_D \eta'_D} \theta_{\eta'_D h_D} \simeq - \frac{2 (m_{q_2} - m_{q_1}) m_{q_2} m_{q_1}}{(m_{q_2}+m_{q_1})^3} \frac{m_{\pi_D}^4}{m_{\eta'_D}^2 m_{h_D}^2} \frac{F_{\pi_D} \theta_D}{v_D} \ ,
\end{eqnarray}
which is proportional to both the sources of CP-invariance violation, $\theta_D$, and ``isospin'' violation, $m_{q_2}-m_{q_1}$. The mixing between the dark pion $\pi^0_D$ and the dark Higgs $h_D$, combined with the Higgs portal interaction proportional to $\kappa$, allows the dark pion to decay directly into SM degrees of freedom, as will be discussed in Section~\ref{sec:darkpiondecay}.

%%%%%%%%%%%%%%%%%%%%%%%%%%%%%%%%%%%%%%%%%%%%%%%%%%%%%%%%%%%%%%%%%%%%%%

\setcounter{equation}{0}
\setcounter{footnote}{0}
\section{Dark Sector Cosmology}
\label{sec:cosmo}

In this section, we discuss the possible role of dark sector particles, the dark leptons (related to neutrino mass) and the dark hadrons (related to dark matter), in the evolution of the early universe, and their possible imprint on cosmological observations.

\subsection{Thermalization in the early universe}

In the early universe, at temperatures much higher than all mass scales in the SM and in the dark sector, all SM and dark sector degrees of freedom can be treated as massless particles, and the two sectors communicate via the Higgs portal interaction governed by the coupling constant $\kappa$, defined in Eq.~(\ref{eq:Ltot}). We require the SM and dark sector to reach thermal equilibrium before the scale of dark $SU(2)$ breaking, $v_D$. Concentrating on $v_D\gg v$, to be justified a posteriori, the process $H_D H_D^{\dagger}\leftrightarrow  HH^{\dagger}$ is in thermal equilibrium if  
\begin{equation}
\kappa\gtrsim\sqrt{\frac{v_D}{M_{\rm Pl}}}\sim 10^{-7}\left(\frac{v_D}{10^3~\rm TeV}\right)^{1/2} \ .
\end{equation} 
At some lower temperature, the SM and dark sectors will decouple. Below the scale $v_D$, we can integrate out the dark sector Higgs scalar and, around the dark $SU(3)$ confinement scale $\Lambda_3$ (assuming it is larger than the electroweak scale $v$), the dark protons interact with the SM Higgs field via the dimension-five effective Lagrangian 
\begin{equation}
\mathcal{L}_{\rm eff} = \frac{y_{p_D} \kappa v_D}{m_{h_D}^2}\bar{p}_Dp_DH^{\dagger}H \ ,
\end{equation}
where $y_{p_D}$ is the effective coupling between the dark proton and the dark Higgs boson. We estimate that the reaction $p_D\bar{p}_D\leftrightarrow HH^{\dagger}$ is no longer in thermal equilibrium around the dark proton mass $m_{p_D}$ if
\begin{equation}\label{kappaupper}
\kappa \lesssim 10^{-1}\left(\frac{10^{-2}}{y_{p_D}}\right)\left(\frac{v_D}{10^3~\rm TeV}\right)\left(\frac{100~\rm TeV}{m_{p_D}}\right)^{3/2} \ .
\end{equation}
We assume $m_{h_D} \simeq v_D$ hereafter, unless otherwise noted. We discuss our estimate for the value of $y_{p_D}$ in Section~\ref{sec:detection}.

In summary, there are values of $\kappa$ such that, at very high temperatures $T\gg v_D$, the SM and dark sector degrees of freedom are in thermal equilibrium and such that, by the time of the dark $SU(3)$ phase transition, the SM and dark sector are thermally and chemically decoupled from one another. Qualitatively, the time evolution of the early universe is as follows:
\begin{itemize}
\item $T\gg  v_D$. The universe is well described as a relativistic gas of SM and dark sector degrees of freedom in thermal equilibrium.
\item $v_D>T>T_{\rm dec}$ ($T_{\rm dec}$ is the temperature related to the thermal decoupling of the SM and dark sectors). The universe is well described as a relativistic gas of SM degrees of freedom, dark quarks and gluons, and left-handed antineutrinos. 
\item As the universe cools down further, but for temperatures above or around the electroweak symmetry breaking scale $v$, three things happen. (i) The dark $SU(3)$ phase transition. At this point, the dark sector degrees of freedom are best described as dark hadrons; (ii) the dark leptons decouple from the rest of the dark sector, and (iii) the SM and dark sectors decouple. 
\item At low enough temperatures, the universe is well described as a relativistic gas of SM particles, a dark lepton gas, and a dark hadron gas --- the three components are no longer in thermal contact with one another. The dark hadrons interact via dark pion exchange. As the universe cools down, the dark baryons ($p_D$ and potentially $n_D$) annihilate into dark pions, until they freeze out. 
\item At lower temperatures still, the dark pions decay. We are left with the SM plus a relic abundance of heavy dark protons. If the dark leptons are stable or very long-lived, they are also present. In the limit where they are very light (e.g., if the neutrinos are Dirac fermions), they contribute to the radiation content of the universe at different epochs.  
\end{itemize}  

\subsection{Very light dark leptons and $\Delta N_{\rm eff}$}

If the mostly-sterile neutrinos $(\nu_D)_{1,2}$ are light enough and relativistic during the time of BBN or CMB formation, we need to worry about their contribution to the effective number of neutrinos, $\Delta N_{\rm eff}$. As discussed above, early in the universe's timeline, the dark sector and the SM are in thermal equilibrium and the dark leptons have the same temperature as the SM neutrinos. After the two sectors decouple, the temperatures of the dark and SM neutrinos will diverge because of the different changes in the degrees of freedom in the two sectors, especially the QCD (dark $SU(3)$) phase transitions when the (dark) quarks and (dark) gluons confine into (dark) hadrons.

If the two sectors decouple at the dark $SU(2)$ breaking scale, which is well before both the SM QCD and dark $SU(3)$ phase transitions, these transitions will heat up their respective neutrinos independently. In this case, the temperature ratio of the SM to dark leptons is
\begin{eqnarray}
\frac{T_\nu}{T_{\nu_D}} = \left( \frac{g^{\rm SM}_{*S}(m_t)}{g^{\rm SM}_{*S}(m_e)} \right)^{1/3} \left( \frac{g^{\rm D}_{*S}(m_{\nu_D})}{g^{\rm D}_{*S}(v_{D})} \right)^{1/3} = \left( \frac{106.75}{10.75} \right)^{1/3} \left( \frac{3.5}{50.5} \right)^{1/3} \simeq 0.88 \ . 
\end{eqnarray}
Under these conditions, the dark leptons are hotter than the SM neutrinos. The contribution of dark leptons to $\Delta N_{\rm eff} = 2/0.88^3 \simeq 2.9$. This is inconsistent with cosmological data, $\Delta N_{\rm eff}=-0.01\pm 0.18$ \cite{Ade:2015xua}.

If the two sectors decouple after the dark $SU(3)$ phase transition happens but before the SM QCD transition, the phase transition in the dark sector will reheat both SM neutrinos and the dark leptons, while the SM QCD phase transition will only reheat the SM neutrinos. In this case, 
\begin{eqnarray}\label{Tratio}
\frac{T_\nu}{T_{\nu_D}} = \left( \frac{g^{\rm SM}_{*S}(m_t)}{g^{\rm SM}_{*S}(m_e)} \right)^{1/3} = \left( \frac{106.75}{10.75} \right)^{1/3} \simeq 2.15 \ ,
\end{eqnarray}
and the contribution of dark leptons to $\Delta N_{\rm eff} = 2/2.15^3 \simeq 0.2$, which is consistent with current data.
Based on Eq.~(\ref{kappaupper}), we find that for the two sectors to still be in equilibrium around the dark QCD phase transition (which is also around the dark proton mass), we need the Higgs portal coupling to be
\begin{eqnarray}
\kappa\gtrsim 0.1 \ .
\end{eqnarray}

One assumption associated with Eq.~(\ref{Tratio}) is that the dark pion mass is not much lighter than the dark $SU(3)$ scale $\Lambda_3$. If the dark pion mass is well below $\Lambda_3$, it might remain in thermal equilibrium with the dark leptons via the dark SU(2) gauge interaction after the SM and dark sectors decouple from one another. Assuming that all species are relativistic through decoupling and that the only relevant process is $\nu_D \overline{\nu}_D \longleftrightarrow \pi_D^\pm \pi_D^\mp$, we estimate that the dark leptons decouple from the dark pions at a temperature $T_{\rm dec}^{\nu_D}\sim 20-25$ GeV, for $v_D = 1$ PeV. While this estimate depends on the total, effective number of relativistic degrees of freedom at decoupling, $g_*$,\footnote{Not to be confused with $g_{*S}$.} this dependence is relatively weak; a more detailed analysis of the degrees of freedom in the SM plasma than the one used in our estimate would not yield a significantly different result. If the dark pion mass is larger than $T_{\rm dec}^{\nu_D}$, dark lepton--dark pion interactions will further reheat the the dark leptons relative to the SM neutrinos in such a way that
\begin{eqnarray}
\frac{T_\nu}{T_{\nu_D}} = \left( \frac{g^{\rm SM}_{*S}(m_t)}{g^{\rm SM}_{*S}(m_e)} \right)^{1/3} \left( \frac{g^{\rm D}_{*S}(m_{\nu_{D}})}{g^{\rm D}_{*S}(m_{\pi_D})} \right)^{1/3} = \left( \frac{106.75}{10.75} \right)^{1/3} \left( \frac{3.5}{6.5} \right)^{1/3} \simeq 1.75 \ .
\end{eqnarray}
This leads to $\Delta N_{\rm eff} = 2/1.75^3 \simeq 0.37$, consistent with cosmological data at around the two sigma level. 
If the dark pions are significantly lighter than $T_{\rm dec}^{\nu_D}$, their relic population is ultimately converted into a non-thermal population of dark leptons once the dark pions decay. We don't explore this possibility any further here but discuss the physics of dark pion decay in Sec.~\ref{sec:darkpiondecay}.

\subsection{The heavy dark lepton window}

The physics of heavier dark leptons is qualitatively different. In particular, as discussed in Sec.~\ref{sec:neutrino}, if a dark lepton is heavier than, roughly, 100\,MeV, it is expected to decay into SM degrees of freedom before BBN. There are virtually no cosmological constraints in this case. On the other hand, according to Fig.~\ref{fig:nulife}, even if the active--sterile mixing angle saturates the current experimental upper bounds, dark leptons with masses below roughly 10\,MeV decay during or after the BBN, injecting non-thermal SM neutrinos as well as photons (with a smaller branching ratio). This possibility is strongly constrained~\cite{Jedamzik:2006xz}.

Another concern is that these long-lived dark leptons may decouple from the thermal bath when they are still relativistic and turn non-relativistic before they decay. Under these circumstances they could temporarily dominate the energy density of the universe and leave an indelible imprint in cosmic surveys. This dark lepton (matter-) dominated universe begins, roughly, when the SM photon temperature drops below the dark lepton mass. The corresponding Hubble time can be roughly estimated to be
\begin{eqnarray}
t_{MD} \sim \frac{M_{pl}}{m_{\nu_D}^2} \simeq 0.1\,{\rm sec}\left( \frac{10\,\rm MeV}{m_{\nu_D}} \right)^2\ .
\end{eqnarray}
On the other hand, for $m_{\nu_D}$ below 10\,MeV the dark leptons mainly decay into three active neutrinos (see Appendix~\ref{sec:SterileNuAppendix} for more details), and the lifetime is given by
\begin{eqnarray}
\tau_{\nu_D} \sim 0.1\,{\rm sec}\left( \frac{10\,\rm MeV}{m_{\nu_D}} \right)^5\left(\frac{1}{|U|^2}\right) \ ,
\end{eqnarray}
where we set all $|U_{\alpha4}|^2,|U_{\alpha5}|^2=|U|^2$, as in Fig.~\ref{fig:nulife}. Clearly, lighter dark leptons live longer and, for $m_{\nu_D}<10\,$MeV, $t_{MD}$ is always much smaller than $\tau_{\nu_D}$, keeping in mind that $|U|^2\ll 1$.

The two problems identified above -- injection of non-thermal neutrinos and photons after BBN and a period of matter domination after BBN --  are alleviated for $m_{\nu_D}$ values below tens of eV. In this case, the mostly-sterile neutrinos serve as a subdominant (hot) dark matter species and their lifetime are longer than the age of the universe. 

In summary, the discussions above point to two distinct mass windows for the dark leptons. If $m_{\nu_D}$ is below dozens of eV, or if the neutrinos are Dirac or pseudo-Dirac fermions, dark leptons behave as dark radiation and contribute to $\Delta N_{\rm eff}$, as discussed in the previous section. If $m_{\nu_D}$ is larger than, roughly, 100\,MeV dark leptons decay quickly enough (within 0.1 second) to avoid bounds from cosmological observables. 
 
\subsection{Dark matter thermal relic abundance}\label{TFO}

The absence of a massless dark gauge boson after spontaneous symmetry breaking and confinement allows for a significant relic abundance for the lightest dark baryons even in the absence of a primordial dark baryon asymmetry. At low enough temperatures, the dark protons interact predominantly with the lightest pseudoscalar dark mesons, including the dark pions $\pi_D^0, \pi_D^\pm$ and the dark eta-prime meson $\eta'_D$. 

\begin{figure}[t]
\centerline{\includegraphics[scale=0.35]{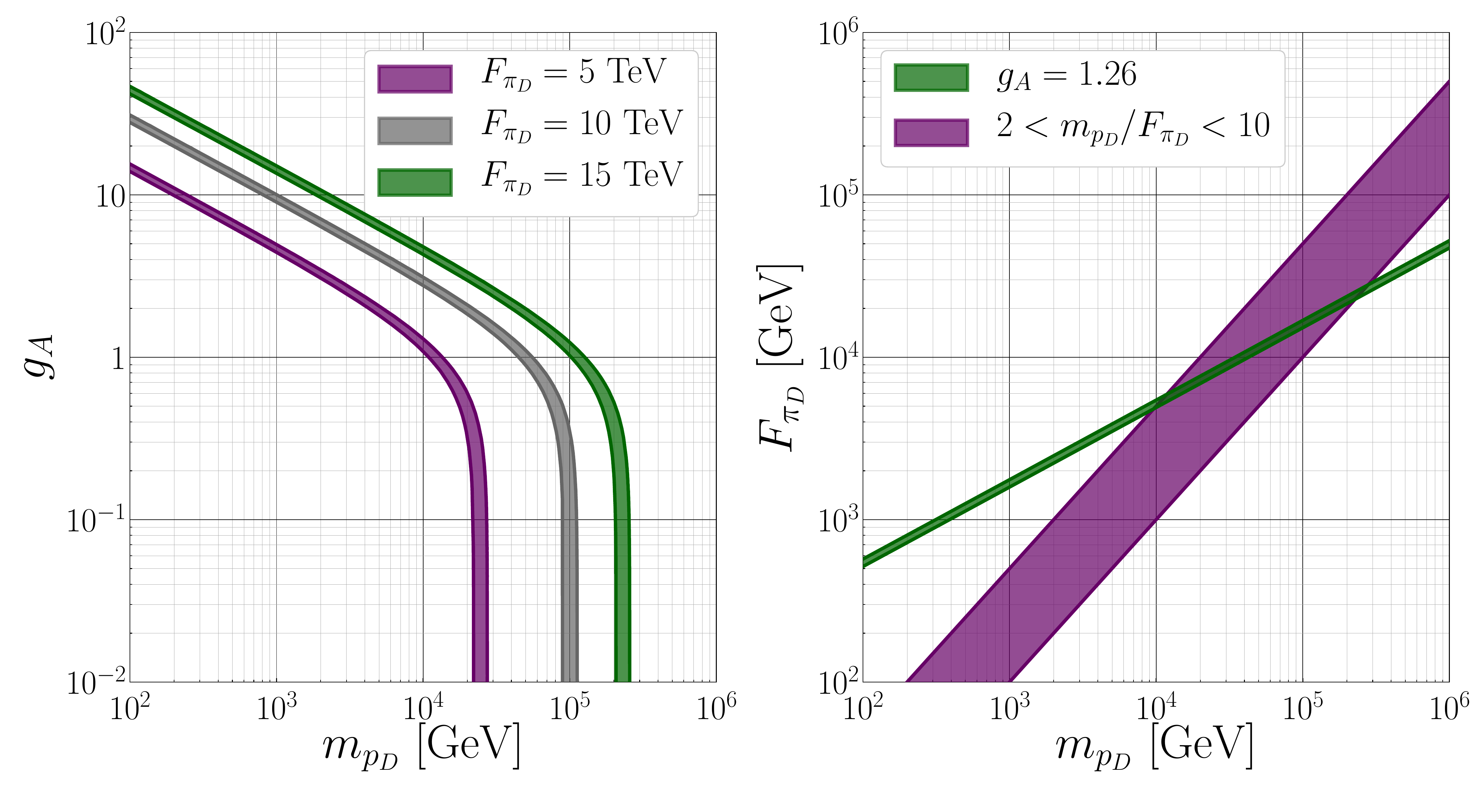}}
\caption{Regions of parameter space as a function of the dark proton mass $m_{p_D}$ and the axial coupling constant $g_A$ (left) or dark pion decay constant $F_{\pi_D}$ (left) for which the relic abundance of dark matter $\Omega_c h^2$ lies between $0.1$ and $0.15$. The measured value of $\Omega_c h^2$ is $0.1199 \pm 0.0022$~\cite{Ade:2015xua}. The left panel depicts the allowed region of parameter space for three different values of $F_{\pi_D}$, 5 TeV (purple), 10 TeV (grey), and 15 TeV (green). The right panel depicts the allowed region (green) for $g_A = 1.26$, consistent with the SM value of the same parameter, and the region (purple) where $m_{p_D}$ is between $2F_{\pi_D}$ and $10 F_{\pi_D}$, again mirroring the relative size of these parameters in the SM.}\label{fig:freezeout}
\end{figure}

Using the Lagrangian in Eq.~(\ref{Lchpt}), the relevant dark nucleon--anti-nucleon annihilation cross-sections, up to terms of order $v_{\rm rel}^2$ in the small relative velocity expansion, are
\begin{eqnarray}\label{Xsecs}
\begin{split}
&(\sigma v_{\rm rel})_{p_D\bar p_D \to \pi_D^0 \pi_D^0} = (\sigma v_{\rm rel})_{n_D\bar n_D \to \pi_D^0 \pi_D^0} =
\frac{g_A^2 m_{p_D}^2}{192 \pi F_{\pi_D}^4} v_{\rm rel}^2 \ , \\
&(\sigma v_{\rm rel})_{p_D\bar p_D \to \pi_D^+ \pi_D^-} = (\sigma v_{\rm rel})_{n_D\bar n_D \to \pi_D^+ \pi_D^-} =
\frac{(g_A^2 + g_V^2)^2 m_{p_D}^2}{64 \pi F_{\pi_D}^4} + \frac{(g_A^2 - g_V^2)^2 m_{p_D}^2}{384 \pi F_{\pi_D}^4} v_{\rm rel}^2 \ ,  \\
&(\sigma v_{\rm rel})_{p_D\bar p_D \to \eta'_D \eta'_D} = (\sigma v_{\rm rel})_{n_D\bar n_D \to \eta'_D \eta'_D} =
\frac{g_A^4 m_{p_D}^2 (32 - 64 r^2 + 48 r^4 -16 r^6 + 3 r^8) \sqrt{1-r^2}}{384 \pi F_{\pi_D}^4 (2 - r^2)^4} v_{\rm rel}^2 \ , \\
&(\sigma v_{\rm rel})_{p_D\bar p_D \to \pi_D^0 \eta'_D} = (\sigma v_{\rm rel})_{n_D\bar n_D \to \pi_D^0 \eta'_D} =
\frac{g_A^4 m_{p_D}^2(1-r^2/4)}{96 \pi F_{\pi_D}^4} v_{\rm rel}^2 \ , \\
&(\sigma v_{\rm rel})_{p_D\bar n_D \to \pi_D^+ \pi_D^0} = (\sigma v_{\rm rel})_{n_D\bar p_D \to \pi_D^- \pi_D^0} =
\frac{(g_A^2 + g_V^2)^2 m_{p_D}^2}{32 \pi F_{\pi_D}^4} 
- \frac{(3g_A^2 - g_V^2)(g_A^2 + g_V^2) m_{p_D}^2}{192 \pi F_{\pi_D}^4} v_{\rm rel}^2 \ , \\
&(\sigma v_{\rm rel})_{p_D\bar n_D \to \pi_D^+ \eta_D'} = (\sigma v_{\rm rel})_{n_D\bar p_D \to \pi_D^- \eta_D'} =
\frac{g_A^4 m_{p_D}^2 (1-r^2/4)}{48 \pi F_{\pi_D}^4} v_{\rm rel}^2 \ .
\end{split}
\end{eqnarray}
Here $m_{p_D}$ is the dark proton mass and $r \equiv m_{\eta_D'}/m_{p_D}$. We set the dark pion mass $m_{\pi_D}$ to zero but allow for a nonzero dark $\eta'_D$ mass $m_{\eta_D'}$.
Because the temperature is still quite high during freeze out, we consider both the dark neutron and dark proton are present in the universe and neglect their mass difference in the results above. 

\begin{figure}[!t]
\centerline{\includegraphics[scale=0.6]{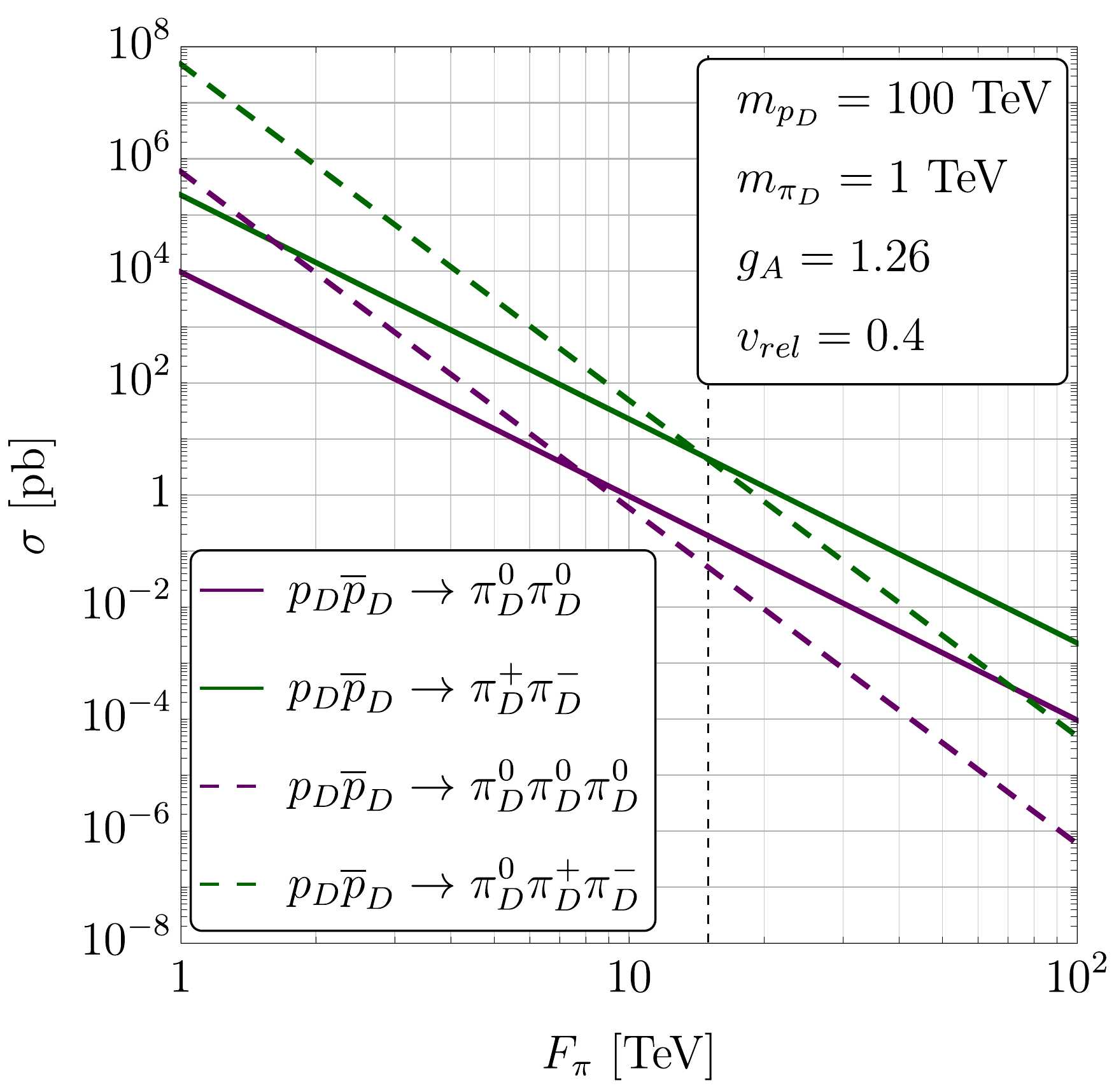}}
\caption{Cross sections for $p_D \overline{p}_D$ annihilation into $\pi_D^0 \pi_D^0$ (solid purple), $\pi_D^+ \pi_D^-$ (solid green), $\pi_D^0 \pi_D^0 \pi_D^0$ (dashed purple) and $\pi_D^0 \pi_D^+ \pi_D^-$ (dashed green) as a function of $F_\pi$. We choose $m_{p_D} = 100$ TeV, $m_{\pi_D} = 1$ TeV and $g_A = 1.26$, and calculate the cross sections for $v_{rel} = 0.4$. The black, dashed line indicates $F_\pi = 15$ TeV; this is our benchmark value in our analyses that follow. We have neglected decays into states containing $\eta^\prime_D$, as these are always subdominant in the parameter space that we explore.}\label{fig:many_pions}
\end{figure}

Based on these Born-level cross-sections\footnote{We assume the dark pion mass is small enough so that the non-perturbative corrections are small. As discussed in Fig.~\ref{fig:SE:S=0} below, Sommerfeld enhancement of the annihilation cross-section could be significant for $m_{\pi_D}$ larger than a few TeV.}, we calculate the thermal freeze out of dark nucleons via annihilation and co-annihilation~\cite{Griest:1990kh}. The average value of $v_{\rm rel}^2$ is $\langle v_{\rm rel}^2\rangle\sim 6T_f/m_{p_D}$, where the freeze-out temperature $T_f\sim m_{p_D}/30$.
The region of parameter space where the dark nucleon relic abundance agrees with the present dark matter abundance is depicted in Fig.~\ref{fig:freezeout}. 
Fig.~\ref{fig:freezeout} (left) depicts the allowed region in the $m_{p_D}$ versus $g_A$ parameter space for different values of $F_{\pi_D}$, while Fig.~\ref{fig:freezeout}(right) depicts the allowed region in the $m_{p_D}$ versus $F_{\pi_D}$ parameter space for fixed $g_A=1.26$.\footnote{This is consistent with the SM QCD value. We also highlight that, as can be seen in Fig.~\ref{fig:freezeout}(left), the relic density is mostly independent of $g_A$ for $g_A\lesssim 0.5$ because the contact interactions $\bar p_D p_D \pi_D \pi_D$ proportional to $g_V^2$ dominate in this case.}  Fig.~\ref{fig:freezeout}(right) also depicts the region of parameter space characterized by $m_{p_D}/F_{\pi_D}\in [2,10]$, qualitatively consistent with what is known about $m_N$ and $F_{\pi}$ in SM QCD. Using SM QCD as guidance, we find the dark matter relic density can be accommodated if the dark proton mass lies between roughly 10--100\,TeV.

For simplicity, we have neglected final states with more than two dark pions. These enhance the overall annihilation cross section and, in turn, point to heavier dark proton masses assuming these are the dark matter. Figure~\ref{fig:many_pions} shows the cross sections for $p_D \overline{p}_D$ annihilation into $\pi_D^0 \pi_D^0$ (solid purple), $\pi_D^+ \pi_D^-$ (solid green), $\pi_D^0 \pi_D^0 \pi_D^0$ (dashed purple) and $\pi_D^0 \pi_D^+ \pi_D^-$ (dashed green) as a function of $F_\pi$. We have chosen $m_{p_D} = 100$ TeV, $m_{\pi_D} = 1$ TeV and $g_A = 1.26$, and have calculated the cross sections for $v_{rel} = 0.4$, corresponding to a temperature close to dark proton freeze-out. For values of $F_\pi \lesssim 15$ TeV (indicated by a black, dashed line), $\pi_D^0 \pi_D^+ \pi_D^-$ is the dominant decay channel; as $F_\pi$ is further decreased, the preferred annihilation channel is into an increasing number of charged dark pions. Above $F_\pi \gtrsim 15$ TeV, $\pi_D^+ \pi_D^-$ is the most important final state; annihilation into many-pion final states is suppressed. Final states containing $\eta^\prime_D$s are always subdominant in the parameter space we explore.

Finally, we comment on the stability of the dark proton as the dark matter candidate. As argued above, it is stable because of a global dark baryon symmetry, $U(1)_{DB}$. However, this global symmetry is anomalous with respect to the $SU(2)$ gauge interaction~\cite{tHooft:1976rip} and therefore only approximately conserved. At zero temperature, the decay of the dark baryon mediated by instantons is exponentially suppressed by $e^{-4\pi/\alpha_2}$. If the dark sector $SU(2)$ gauge coupling is modestly small ($\alpha_2 \lesssim0.1$), the dark proton lifetime is much larger than the age of the universe ($\tau_{p_D}>10^{26}\,$sec), safely beyond constraints on decaying dark matter. 

\subsection{Dark pion decay}\label{sec:darkpiondecay}

Immediately after the dark nucleons freeze out, the pseudoscalar dark mesons are still present in the universe. They are, however, unstable particles. If heavy enough, the dark $\eta_D'$ will dominantly decay, promptly, into dark pions while the dark pions can only decay ``weakly,'' and only if the decays are kinematically allowed.  Here we discuss the three potentially relevant dark pion decay-modes. Throughout, we assume the dark pions are heavier than the dark leptons, even if those are allowed Majorana masses. 

Dark pions can decay into two dark leptons, similar to $\pi^\pm$ decays in the SM. Here the ``neutral'' pions also decay in the same way, via $X_3$ exchange (there are no light ``dark photons'' into which the $\pi_D^0$ can decay). The decay rate is 
\begin{eqnarray}
\Gamma_{\pi_D\to2\nu_D} = 2\frac{G_{F_D}^2 F_{\pi_D}^2 m_{\pi_D} m_{\nu_D}^2}{4\pi} \ ,
\end{eqnarray}
where $G_{F_D} = 1/(\sqrt{2} v_D^2)$ is the dark-sector analog of the Fermi constant and we assume the two dark leptons have the same mass. As is the case of SM pion decay, this decay rate is helicity suppressed, proportional to the dark lepton masses. For very light dark leptons, or if the neutrinos are Dirac fermions, this decay rate can be exceptionally small.

Dark pions can decay into four neutrinos through the emission of two virtual dark $SU(2)$ bosons. This decay rate is not helicity suppressed, and can be estimated as
\begin{eqnarray}
\Gamma_{\pi_D\to4\nu_D} = \frac{1}{2m_{\pi_D}} \int d\Pi_f^{\rm 4\, body} |\mathcal{M}|^2 \sim \frac{1}{2m_{\pi_D}} \left( \frac{m_{\pi_D}^4}{24576\pi^5} \right) \left(\frac{g_2^8 F_{\pi_D}^2 m_{\pi_D}^4}{M_X^8} \right)
\sim \frac{G_{F_D}^4 F_{\pi_D}^2 m_{\pi_D}^7}{48 \pi^5} \ ,
\end{eqnarray}
where we assumed $|\mathcal{M}|^2$ has no final state momentum dependence, and made use of the volume of the massless $n$-body final state phase space, $\int d\Pi_f^{n body} =\left[4(2\pi)\Gamma(n) \Gamma(n-1) \right]^{-1} \left[ {s}/({16\pi^2}) \right]^{n-2}$. In the second bracket of the above equation, the decay matrix element square is estimated using naive dimensional analysis.

\begin{figure}[t]
\centerline{\includegraphics[scale=0.35]{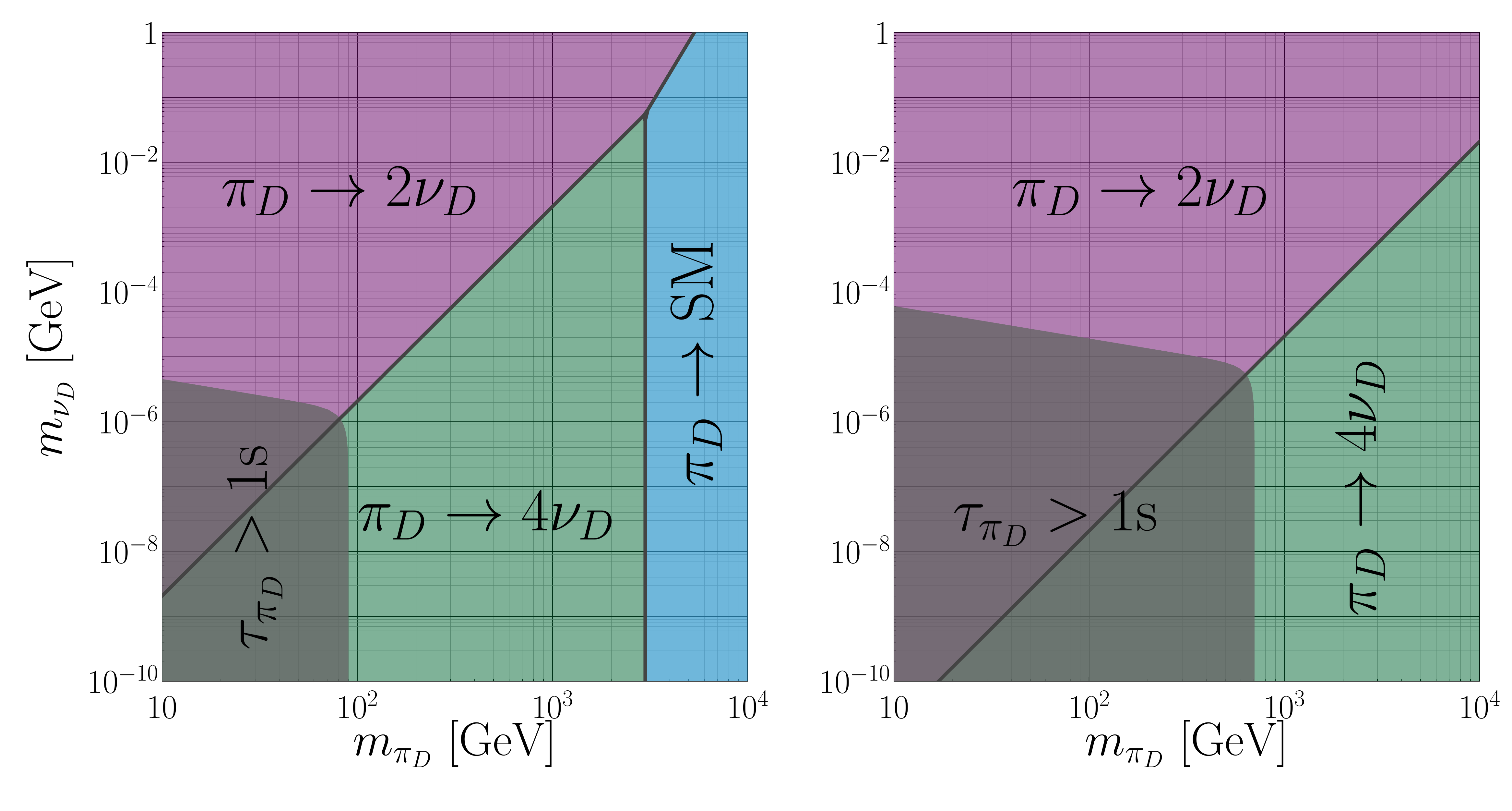}}
\caption{The dominant channel for dark pion $\pi_D$ decay as a function of the dark pion mass, $m_{\pi_D}$ and dark neutrino mass $m_{\nu_D}$. Purple regions denote where the decay channel $\pi_D \to 2\nu_D$ is preferred, green regions denote where $\pi_D\to 4\nu_D$ is preferred, and blue denotes where decays to the standard model, largely $\pi_D \to t\bar{t}$, are preferred. In both figures, $\kappa$ and $\theta_D$ are set to $0.1$. In the left figure, $v_D = 2m_{h_D} = 100$ TeV, $m_{\eta'_D} = 7$ TeV, and $F_{\pi_D} = 2$ TeV, and in the right figure $v_D = 2m_{h_D} = 1$ PeV, $m_{\eta'_D} = 70$ TeV, and $F_{\pi_D} = 15$ TeV. We choose these parameters to keep the dark proton mass equal to $v_D/10$, and the ratio between the dark eta prime meson and the proton equal to $0.7$. We choose $F_{\pi_D}$ to satisfy the observed dark matter relic abundance, as depicted in Fig.~\ref{fig:freezeout}.}\label{fig:piondecay}
\end{figure}

The presence of a nonzero  dark $\theta_D$-term allows the pseudoscalar dark mesons to mix with the dark Higgs boson (see Eq.~(\ref{darkpionHiggs})) and, via the $\kappa$-mediated interaction, mix with the SM Higgs boson. This leads to decays of the dark pions directly into SM degrees of freedom.  For dark pions heavier than the SM electroweak scale, this decay rate is 
\begin{eqnarray}
\Gamma_{\pi_D\to {\rm SM}} &=& \theta_{\pi^0 h}^2 \left[ \frac{3 m_t^2 m_{\pi_D}}{8 \pi v^2} \left( 1 - \frac{4m_t^2}{m_{\pi_D}^2} \right)^{3/2} \Theta(m_{\pi_D} - 2 m_t) + (t\to b) \right. \nonumber \\ 
& & \left. + \frac{m_{\pi_D}^3}{16\pi v^2}\left( 1 - \frac{4 M_W^2}{m_{\pi_D}^2} + \frac{12 M_W^4}{m_{\pi_D}^4} \right) \sqrt{1 - \frac{4 M_W^2}{m_{\pi_D}^2}}\Theta(m_{\pi_D} - 2 M_W) + (W\to Z) 
\right] \ ,
\end{eqnarray}
where
\begin{eqnarray}\label{eq:thetapih}
\theta_{\pi^0_D h} \sim \frac{m_{\pi_D}^4}{m_{\eta_D'}^2 m_{h_D}^2} \frac{F_{\pi_D} \theta_D}{v_D} \times \frac{\kappa v v_D}{m_{h_D}^2} \ .
\end{eqnarray}

Fig.~\ref{fig:piondecay} depicts the regions of the $m_{\pi_D}$ versus $m_{\nu_D}$ parameter space where one of the three dark pion decay modes discussed above dominates. 
In the region shaded in gray, the dark pion lifetime is long enough that, in the early universe, relic dark pions decay during or after BBN, a condition that might be challenged by cosmological data~\cite{Jedamzik:2006xz}.
For light enough dark leptons, an upper bound on the dark pion lifetime translates into a lower bound on the dark pion mass. For large $\theta_D$ values and light dark Higgs masses, the direct decay of dark pions into SM degrees-of-freedom can be dominant (see Fig.~\ref{fig:piondecay} (left)). 

%%%%%%%%%%%%%%%%%%%%%%%%%%%%%%%%%%%%%%%%%%%%%%%%%%%%%%%%%%%%%%%%%%%%%%

\setcounter{footnote}{0}
\setcounter{equation}{0}
\section{Expectations for Dark Matter Searches}
\label{sec:detection}

In the previous section, we argued that the lightest dark baryon state in the $SU(3)\times SU(2)$ dark sector model -- the dark proton -- is a plausible dark matter candidate assuming its mass is around tens of TeV. In this section, we discuss the potential for observing such a dark matter candidate, from underground experiments to astrophysical observatories. Current and next-generation indirect-detection experiments turn out to be best positioned to test the hypothesis that the dark matter consists of heavy dark protons. In particular, we calculate in detail the Sommerfeld enhancement of low-velocity dark matter (the dark protons) annihilation due to the exchange of a light pseudoscalar mediator (the dark pions). We expect these detailed results to be useful for phenomenological studies of scenarios beside the model under investigation here.

\subsection{Direct detection}\label{sec:dd}

For the direct detection of the dark proton dark matter, the relevant low energy effective interaction of it with the SM sector goes through the Higgs portal
\begin{eqnarray}
\mathcal{L}_{\rm eff} \supset \lambda_{p_D} \bar p_D p_D h  \ ,
\end{eqnarray}
where $\lambda_{p_D} = {y_{p_D} \kappa v v_D}/({2m_{h_D}^2})$. This is derived from the $H-H_D$ mixing term assuming the dark Higgs is much heavier than the SM Higgs boson. The effective coupling between the dark proton and the dark Higgs is
\begin{eqnarray}\label{darkf}
y_{p_D} =\frac{1}{v_D} \left \langle p_D \left| m_{q_1} \bar q_1 q_1 +  m_{q_2} \bar q_2 q_2  \right| p_D \right\rangle = \frac{m_{p_D}}{v_D} \left(f_{T_1}^D + f_{T_2}^D\right) \ ,
\end{eqnarray}
where $f_{T_1}^D,f_{T_2}^D$ are dimensionless constants which can be estimated using lattice techniques. The spin-independent direct-detection cross-section on a nucleon target is, therefore~\cite{Wise:2014jva},
\begin{eqnarray}
\sigma_{p_D N}^{\rm SI} \simeq \frac{\lambda_{p_D}^2 f^2 m_N^4}{\pi v^2 m_h^4} \ ,
\end{eqnarray}
where we have assumed the DM mass is much larger than the target nucleon mass, $m_N$, and the parameter $f$ is defined in analogy to Eq.~(\ref{darkf}),
\begin{eqnarray}
m_N f = \sum_q \left \langle N \left| m_q \bar q q  \right| N \right\rangle \ .
\end{eqnarray}
According to recent lattice calculations, $f\simeq 0.35$~\cite{Giedt:2009mr}. Experimental results agree with this value of $f$ (see, e.g., Refs.~\cite{Alarcon:2011zs,Alarcon:2012nr}).

The thermal relic abundance of dark protons coincides with dark matter observations for dark proton masses above 10~TeV and, as briefly discussed above, the dominant mechanism for dark matter--regular matter scattering is Higgs exchange. We expect, therefore, tiny direct detection signals. Indeed, for $v_D = m_{h_D}=1$\,PeV, $m_{p_D}=100\,$TeV, $\kappa=0.1$, and $f_{T_u}^D + f_{T_d}^D =0.05$~\cite{Giedt:2009mr}, the direct detection cross-section is
$\sigma_{p_D N}^{\rm SI}\simeq 3\times 10^{-57}\,{\rm cm^2}$, well below current and near-future direct detection limits, and the neutrino floor~\cite{Szydagis:2016few}. It is important to highlight that if, for example, there is a large asymmetry between dark protons and antiprotons, i.e., if the dark matter is asymmetric, a lighter dark sector would preferred. In this case, one expects significantly larger cross-sections.

\subsection{Collider physics}\label{sec:collider}

If light enough, dark pions can be resonantly produced at proton-proton colliders such as the Large Hadron Collider through the mixing mechanism between the SM Higgs boson and the dark pion discussed in Section~\ref{sec:darkpiondecay}. If the dark pion has mass $m_{\pi_D} \sim 1$ TeV, dark pions will decay (promptly) into two  (for heavy $\nu_D$) or four dark leptons (for light $\nu_D$), as depicted in Fig.~\ref{fig:piondecay}.  Heavier dark neutrinos ($m_{\nu_D} \sim 100 - 500$ MeV), in turn, decay preferentially via $\nu_D \to \pi^0 \nu$, $\nu_D \to \pi^\pm e^\mp$, and $\nu_D \to \pi^\pm \mu^\mp$ (see Section~\ref{sec:neutrino} and Appendix~\ref{sec:SterileNuAppendix}). Additionally, these decays occur with lifetimes between $10^{-10} - 10^{-1}$ s, depending on the assumptions regarding the mixing angles $|U_{\alpha4}|^2,|U_{\alpha5}|^2$ (see Fig.~\ref{fig:nulife}). For these dark pion and dark lepton masses, the main signature at a collider would be two displaced vertices of charged leptons and pions, a relatively background-free search.

We roughly estimate the production cross-section of dark pions as 
\begin{align}
\sigma(p p \to \pi_D) &\simeq \sigma(p p \to h)(m_h) \theta_{\pi_D h}^2,
\end{align} 
where $\theta_{\pi_D h}$ is the dark pion--SM Higgs mixing parameter discussed in Section~\ref{sec:darkpiondecay}. The LHC Higgs cross-section working group estimates that the cross-section $\sigma(p p \to h)$ is $\mathcal{O}(10^{-1})$ pb at $m_h \sim 1$ TeV, the mass scale we are interested in for the dark pions~\cite{LHCTwiki}. Using the expression in Eq.~(\ref{eq:thetapih}), for the parameter-values of interest this cross-section is tiny, on the order of $10^{-24} (\theta_D \kappa)^2$ pb. For a lighter dark sector with relatively heavier dark pions or increased $CP$-violation via the $\theta_D$ parameter, on the other hand, the value of $\theta_{\pi_D h}$ is not as small, and collider experiments may provide some sensitivity to this type of new phenomenon. We refer readers to Ref.~\cite{Nemevsek:2016enw} for a phenomenological study of similar collider signals based on a different context.

\subsection{Indirect detection}\label{sec:DMID}

Since the dark matter candidate considered in this work is ``symmetric,'' composed of equal amounts of dark protons and antiprotons,\footnote{Here, for simplicity, we assume the dark neutron is heavy enough compared to the dark proton and decays away quickly, and thus does not play a role in the dark-matter annihilation processes today.} indirect detection signals from dark matter annihilation in, {\it e.g.}, the center of the galaxy are expected. The $p_D$ and $\bar p_D$ particles annihilate predominantly into dark pions and dark eta-primes. The $\eta'_D$ decays promptly into dark pions while the dark pions further decay into dark leptons or directly into SM degrees of freedom, as discussed in Sec.~\ref{sec:darkpiondecay}. If the dark pions decay directly into SM degrees of freedom, indirect detection searches mirror standard indirect detection searches associated to very heavy dark matter particles~\cite{Gunn:1978gr}. If the dark pions decay into dark leptons, the associated indirect detection signals will depend on the lifetime and mass of the dark leptons. 

\begin{itemize}
\item Heavy dark leptons decay quickly into, ultimately, charged leptons, mostly-active neutrinos, or photons.  Since the preferred mass of our dark matter candidate is around tens of TeV, observatories like Fermi-LAT, HESS, HAWC, which are sensitive to multi-TeV gamma rays, are especially suitable.
The current upper limit on the annihilation cross section is roughly $10^{-24}\,{\rm cm^3/s}$~\cite{Lefranc:2016srp,Albert:2017vtb} for dark matter mass around 100\,TeV.
For recent phenomenological studies of indirect detection via the heavy Majorana neutrino portal, see Refs.~\cite{Campos:2017odj,Batell:2017rol}. Neutrino telescopes may be sensitive to the daughter active neutrinos. 

\item Very light dark leptons are cosmologically stable but could manifest themselves in neutrino telescopes~\cite{Aartsen:2015xej, Adrian-Martinez:2015wey} such as IceCube, ANTARES and SuperK. Interactions rates will depend on the magnitude of the active--sterile mixing angle. 
If the mixing angle is order one, the current upper limit on the annihilation cross section is also around $10^{-24}\,{\rm cm^3/s}$ for 100\,TeV dark matter~\cite{Albert:2016emp}.

\item If the neutrinos are Dirac fermions, the dark leptons play the role of right-handed neutrinos. In this case, dark pions decay into (four) SM-gauge singlet fermions with ultra high energies. These in turn, are virtually unobservable.

\end{itemize}

The Born-level annihilation cross section of dark protons and antiprotons is roughly $3\times 10^{-26}\,{\rm cm^3/s}$, as dictated by the dark matter relic abundance.
Dark protons, however, self-interact through dark pion exchange, and we are interested in the case $m_{\pi_D} < m_{p_D}$.  
Hence, today, when the typical dark matter velocities are much lower than those during freeze out, the annihilation cross-section is expected to be enhanced due to non-perturbative effects. In the next subsection, we discuss the general Sommerfeld enhancement associated with light pseudoscalar exchange.

\subsection{Sommerfeld enhancement with a pseudoscalar dark force}

Here, we derive the Sommerfeld enhancement factor associated with the exchange of a light pseudoscalar. For low-velocity dark matter annihilation, Sommerfeld enhancement plays a crucial role in potentially raising the present DM annihilation rate at the center of galaxy to within reach of the current and near-future indirect detection searches. 

The low energy self-interaction between the dark proton and dark antiproton is generated from one dark pion exchange. The relevant interaction terms from Eq.~(\ref{Lchpt}) can be written as
\begin{eqnarray}\label{LeffPINN}
\mathcal{L} \supset \frac{g_A}{2 F_{\pi_D}} \bar N \gamma^\mu \partial_\mu \vec \pi_D \cdot \vec\tau \gamma_5 N = \frac{ig_A m_{p_D}}{F_{\pi_D}} \bar N \vec \pi_D \cdot \vec\tau \gamma_5 N = - \frac{ig_A m_{p_D}}{F_{\pi_D}} \overline{N^c} \vec \pi_D \cdot \vec\tau \gamma_5 N^c \ ,
\end{eqnarray}
where we first work in the isospin conserving limit, $N = (p_D, n_D)^T$, and $\vec\tau$ are the Pauli matrices in isospin space. In the second step we made use of the equations of motion, and in the third step we rewrite the same interaction term for dark antinucleons, $N^c = (n_D^c, -p_D^c)$. This is convenient for deriving the $N\bar N$ potential. The superscript ${^c}$ stands for the charge conjugation of a fermion field. 

In Eq.~(\ref{LeffPINN}), there is a relative minus sign between $\pi_D NN$ and $\pi_D N^c N^c$ interactions~\cite{Lee:1956sw} so the sign of the potential between $N N^c$ is opposite to that between $NN$. The above Lagrangian leads to the following $N N^c$ potential (in momentum space), in the isospin conserving limit:
\begin{eqnarray}\label{VNNC}
V_{N N^c}(\vec q) = \frac{g_A^2}{4 F_{\pi_D}^2} \frac{(\vec \sigma_1 \cdot \vec q)(\vec \sigma_2 \cdot \vec q)}{|\vec q|^2 + m_{\pi_D}^2} (\vec \tau_1 \cdot \vec \tau_2) \ ,
\end{eqnarray}
where $\vec\tau_{1,2}$ are the isospin vectors of $N, N^c$ respectively.

We assume a large enough mass difference between the dark proton and the dark neutron so that today only the dark protons $p_D$ (and dark antiprotons, $\bar p_D$) are around, {\it i.e.}, nature provides an initial state that explicitly breaks the isospin symmetry. The relevant interaction between $p_D, \bar p_D$ goes through one $\pi_D^0$ exchange and takes the form
\begin{eqnarray}
\mathcal{L} = \frac{ig_A m_{p_D}}{F_{\pi_D}} \bar p_D \gamma_5 p_D \pi_D^0 = \frac{ig_A m_{p_D}}{F_{\pi_D}} \overline{p^c_D} \gamma_5 p^c_D \pi_D^0 \ ,
\end{eqnarray} 
which is part of Eq.~(\ref{LeffPINN}). In practice, we simply pick out the $\tau_1^z \tau_2^z$ from the isospin operator in Eq.~(\ref{VNNC}), and find the matrix element between the initial and final states; $\langle p_D \bar p_D|\tau_1^z \tau_2^z|p_D \bar p_D\rangle = -1$. Therefore, the effective interaction between $p_D$ and $\bar p_D$ is
\begin{eqnarray}
V(\vec q) = - \frac{g_A^2}{4 F_{\pi_D}^2} \frac{(\vec \sigma_1 \cdot \vec q)(\vec \sigma_2 \cdot \vec q)}{|\vec q|^2 + m_{\pi_D}^2} \ .
\end{eqnarray}
In position space, the potential energy takes the form
\begin{eqnarray}\label{PotN}
\begin{split}
V(r) &= \frac{g_A^2 m_{\pi_D}^2}{48 \pi F_{\pi_D}^2} \left[ (\vec{\sigma}_1 \cdot \vec{\sigma}_2) V_C(r) + \left( 3 (\vec{\sigma}_1 \cdot \hat{r})(\vec{\sigma}_2 \cdot \hat{r}) -  (\vec{\sigma}_1 \cdot \vec{\sigma}_2) \rule{0mm}{3.5mm}\right) V_T(r) \rule{0mm}{4.5mm}\right] \ , \\
&= \frac{g_A^2 m_{\pi_D}^2}{48 \pi F_{\pi_D}^2} \left[ 2 \left( S(S+1) - \frac{3}{2} \right) V_C(r) + 2 \left( 3 (\vec{S} \cdot \hat{r})^2 -  S(S+1) \rule{0mm}{3.5mm}\right) V_T(r) \right] \ ,
\end{split}
\end{eqnarray}
where $\vec{S}$ is the total spin of the $p_D - \bar{p}_D$ system, and
\begin{eqnarray}\label{VCVT}
V_C(r) = \frac{e^{- m_{\pi_D} r}}{r}, \ \ \ \ \
V_T(r) = \left( 1 + \frac{3}{m_{\pi_D} r} + \frac{3}{m_{\pi_D}^2 r^2} \right)\frac{e^{- m_{\pi_D} r}}{r} \ .
\end{eqnarray}
The interactions above conserve total angular momentum $\vec{J} = \vec{L} + \vec{S}$ and spin $|\vec{S}|^2\to S(S+1)$, but allow $L$ to change by two units through the $V_T$ term (the ``direction'' of $\vec{S}$ changes accordingly). In the galaxy today, the dark matter particles are non-relativistic and most likely to interact in the $s$-wave. Therefore, we will henceforth focus on the $L=0$ state, which may be coupled to the $L=2$ state. The total spin, on the other hand, can be either $S=0$ (singlet state) or $S=1$ (triplet state). We discuss these in turn.

\begin{figure}[t!]
\centerline{\includegraphics[width=0.5\linewidth]{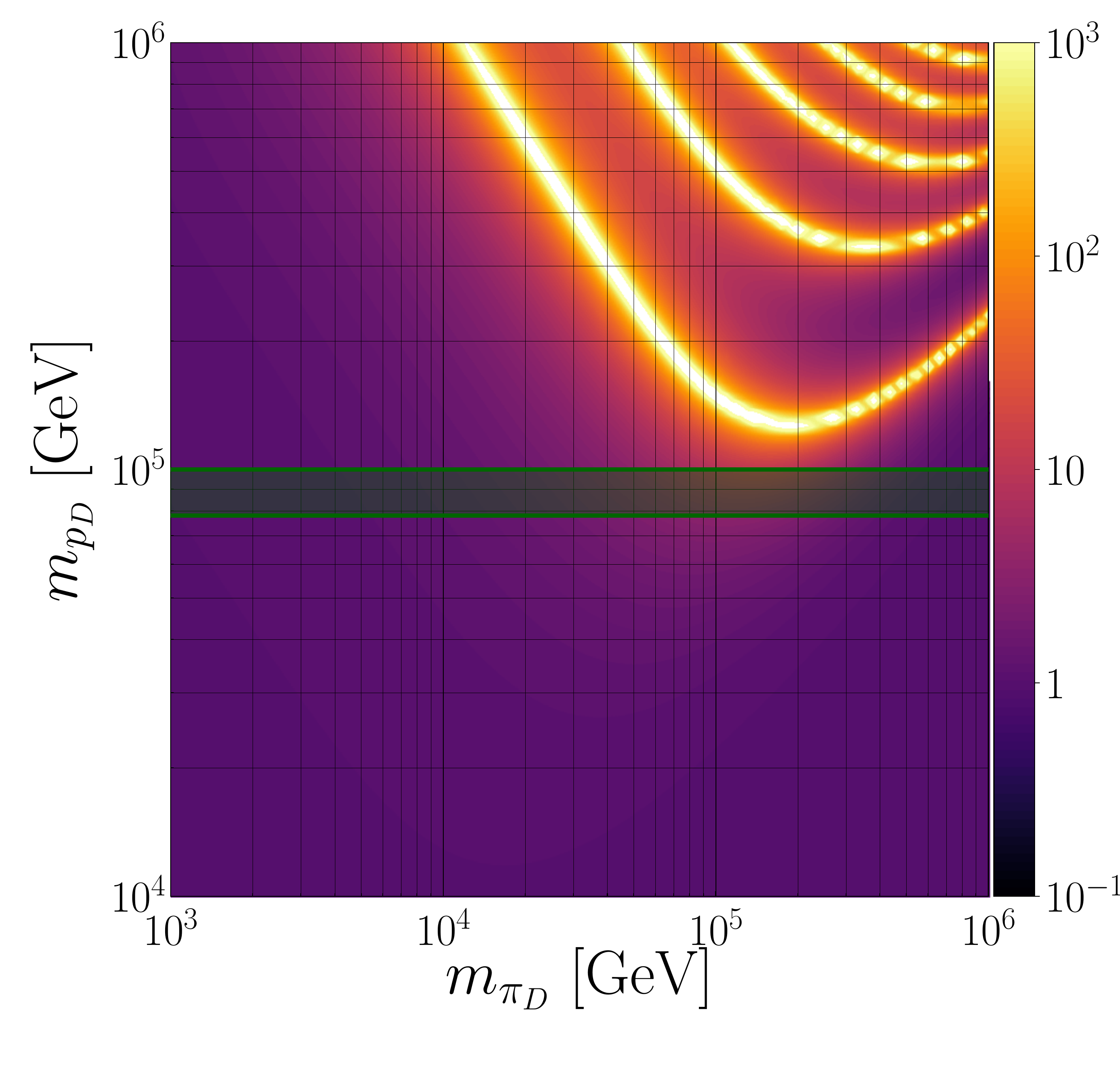}
\includegraphics[width=0.5\linewidth]{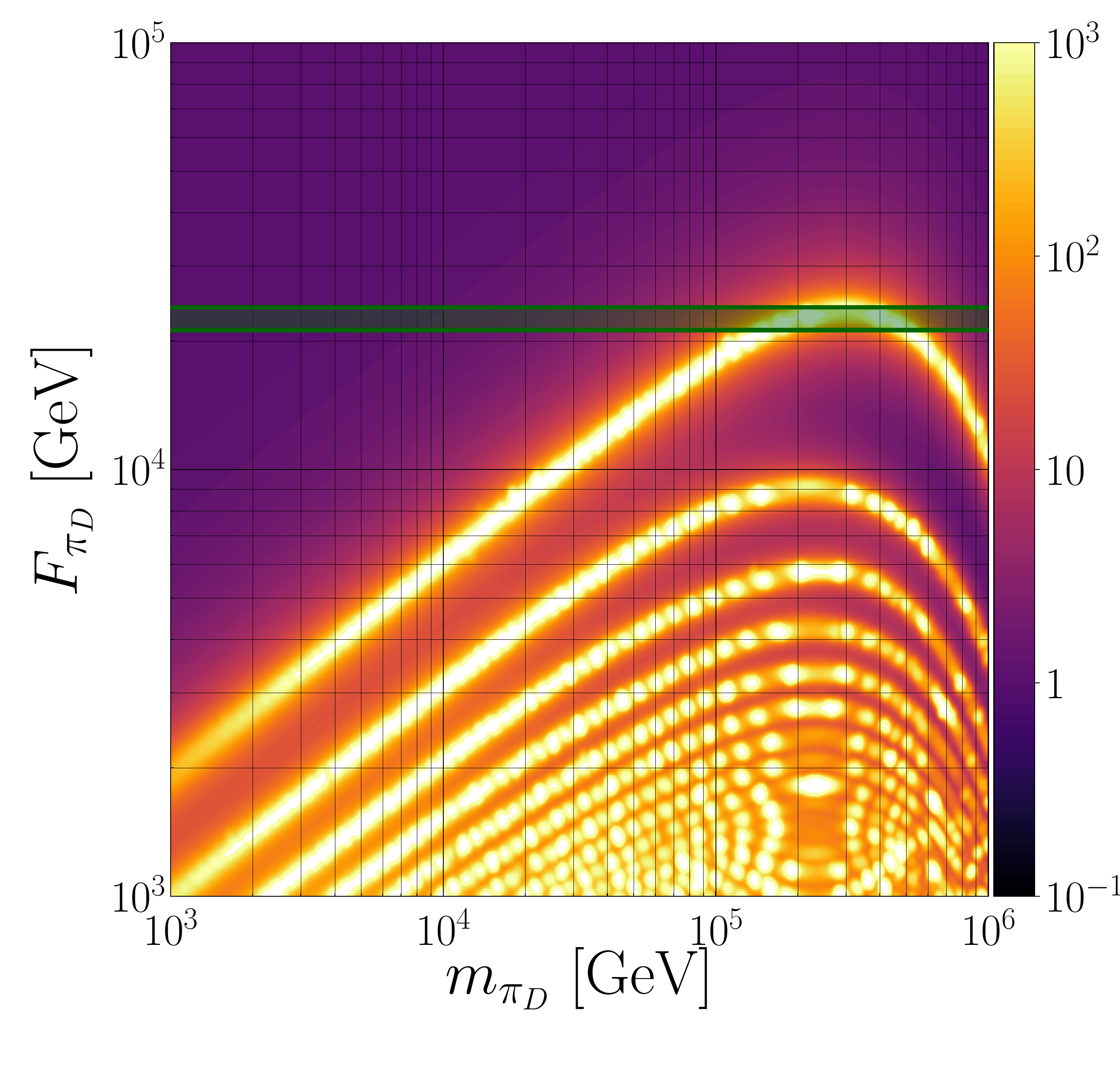}}
\caption{Sommerfeld enhancement of $p_D \bar{p}_D$ annihilation in the $S = 0$, $L = 0$ state as a function of the dark pion mass $m_{\pi_D}$ and dark proton mass $m_{p_D}$ (left) or dark pion decay constant $F_{\pi_D}$ (right). In both plots, $g_A = 1.26$ and $v_\text{rel} = 10^{-3}$, corresponding to the relative velocity of dark matter in the galactic halo at present day. In the left plot, $F_{\pi_D} = 15$ TeV, and, in the right plot, $m_{p_D} = 200$ TeV. Both plots also depict the region (green band) for which the relic density of dark protons agrees with the measured density of dark matter in the universe, per Fig.~\ref{fig:freezeout}.}\label{fig:SE:S=0}
\end{figure}

\bigskip
If $S=0$ and $L=0$, the total angular momentum is $J=0$. In this case, the $L=0, 2$ states do not couple since the $L=2$ state corresponds to a different total angular momentum state, $J=2$. As a result, only the $V_C(r)$ term above can play a role and, effectively, we find
\begin{eqnarray}
V(r) = - \frac{g_A^2 m_{\pi_D}^2}{16 \pi F_{\pi_D}^2} \frac{e^{- m_{\pi_D} r}}{r} \ .
\label{eq:VS0L0}
\end{eqnarray}
This is an attractive Yukawa potential.\footnote{The sign of this potential is opposite to the one found in an earlier study~\cite{Bedaque:2009ri}. A SM analog is the $J/\psi \to p \bar p \gamma$ decay, when the final state $p \bar p$ are near threshold. That interaction is known to be attractive~\cite{Zou:2003zn}.  
} 
The corresponding $s$-wave Sommerfeld enhancement factor for this channel is qualitatively similar to those obtained in the case of massive-vector exchange or real-scalar dark exchange, discussed in Refs.~\cite{Hisano:2004ds, ArkaniHamed:2008qn, Cassel:2009wt}.

The coupling strength in Eq.~(\ref{eq:VS0L0}), $g_A^2 m_{\pi_D}^2/(16 \pi F_{\pi_D}^2)$,  is proportional to the dark pion mass. Therefore, as the dark pion mass goes to zero, the Sommerfeld enhancement disappears, {\it i.e.}, approaches unity. The Sommerfeld enhancement factor as a function of the dark pion mass, in this case, is depicted in Fig.~\ref{fig:SE:S=0}. Fig.~\ref{fig:SE:S=0}(left) depicts the enhancement factor as a function of the dark proton mass $m_{p_D}$, while Fig.~\ref{fig:SE:S=0}(right) depicts the enhancement as a function of the decay constant $F_{\pi_D}$. The behavior observed in Fig.~\ref{fig:SE:S=0}(right) may also be obtained by varying $g_A$ since the potential in Eq.~(\ref{eq:VS0L0}) only depends on the ratio $g_A/F_{\pi_D}$.

\bigskip
If $S=1$, the $L=0, J=1$ state of interest will couple to the $L=2, J=1$ state. The effective potential has a similar form to the deuteron potential~\cite{Gartenhaus:1955zz, Forest:1999wu}, with opposite sign,
\begin{eqnarray}\label{eq:Vpot}
V(r) = \frac{g_A^2 m_{\pi_D}^2}{48 \pi F_{\pi_D}^2} \left[ V_C(r) + 2 \left( 3 (\vec{S} \cdot \hat{r})^2 -  2 \rule{0mm}{3.5mm}\right) V_T(r) \right] \ .
\end{eqnarray}
In this case, we must solve the coupled eigenvalue problem. We define the two states according to their $|LSJM_J\rangle$ quantum numbers. The $M_J=0$ states can be decomposed into orbital angular momentum eigenstates (spherical harmonic functions $Y_{L,M_L}$ in the position-eigenstate basis) and the total spin eigenstates ($|S M_S\rangle$)
\begin{eqnarray}\label{eq:CG}
| 0110 \rangle = Y_{0,0}(\hat r) |10 \rangle, \ \ \ | 2110 \rangle = \sqrt{\frac{3}{10}} Y_{2,-1}(\hat r) |11\rangle - \sqrt{\frac{2}{5}} Y_{2,0}(\hat r) |10\rangle + \sqrt{\frac{3}{10}} Y_{2,1}(\hat r) |1-\!\!1\rangle \ ,
\end{eqnarray}
and the relevant wavefunction in the scattering problem can be written as
\begin{eqnarray}\label{wave}
\Psi_{\vec{k}} (\vec{r}) = R_{0k}(r) | 0110 \rangle + R_{2k}(r) | 2110 \rangle \ .
\end{eqnarray}
In order to derive the equations for $R_{0k}(r)$ and $R_{2k}(r)$, we project the eigenstates $\left|\Psi_{\vec{k}} \right\rangle$ onto the $Y_{00}(\hat r) |10 \rangle$ and $Y_{20}(\hat r) |10 \rangle$ subspaces. This leads to
\begin{equation}\label{CoupledSchrodinger}
\begin{split}
\left[ \frac{1}{2\mu} \frac{1}{r^2} \frac{\partial}{\partial r} \left( r^2 \frac{\partial}{\partial r}\right) + \frac{k^2}{2\mu} - A V_C(r) \right] R_{0k}(r) = \sqrt{8} A V_T(r) R_{2k}(r) \ , \\
\left[ \frac{1}{2\mu} \frac{1}{r^2} \frac{\partial}{\partial r} \left( r^2 \frac{\partial}{\partial r} - \frac{6}{r^2}\right) + \frac{k^2}{2\mu} + 2A V_T(r) - A V_C(r) \right] R_{2k}(r) = \sqrt{8} A V_T(r) R_{0k}(r) \ .
\end{split}
\end{equation}
where $\mu \equiv m_{p_D}/2$ is the reduced mass, and $A \equiv {g_A^2 m_{\pi_D}^2}/({48\pi F_{\pi_D}^2})$.
See Appendix~\ref{appSch} for the details of this derivation.

The $s$-wave Sommerfeld enhancement factor is defined as
\begin{eqnarray}
S_{L=0} = \frac{|R_{0k}(r\to0)|^2}{4\pi} \ ,
\end{eqnarray}
and hence very sensitive to the details of the interaction near the origin. The potential term $V_T(r)$, however, diverges as $1/r^3$ so the wavefunction is not well defined near the origin (its derivative diverges). The reason behind this is that the effective potential Eq.~(\ref{PotN}) is derived at tree-level and in the non-relativistic limit, when the momentum transfer is assumed to be small compared to $4\pi F_{\pi_D}$. At very short distances, multiple $\pi_D$ exchange diagrams become important, and strong corrections to the potential Eq.~(\ref{PotN}) are expected (the theory defined in Eq.~(\ref{LeffPINN}) should contain all ingredients necessary to properly compute the effect from first principles). Moreover, the dark proton is not an elementary particle, its size roughly given by the inverse of the dark confinement scale, of order $m_{p_D}^{-1}$ or $F_{\pi_D}^{-1}$. The effective description using the dark proton as a point-like degree of freedom Eq.~(\ref{PotN}) is expected to be valid only at large distances, for $r$ values larger than some cutoff $r_{p_D}$. For both these reasons, a regularization of the potential is required~\cite{Lepage:1997cs, Beane:2000wh, Bellazzini:2013foa}. 

\begin{figure}[t!]
\centerline{\includegraphics[width=0.5\linewidth]{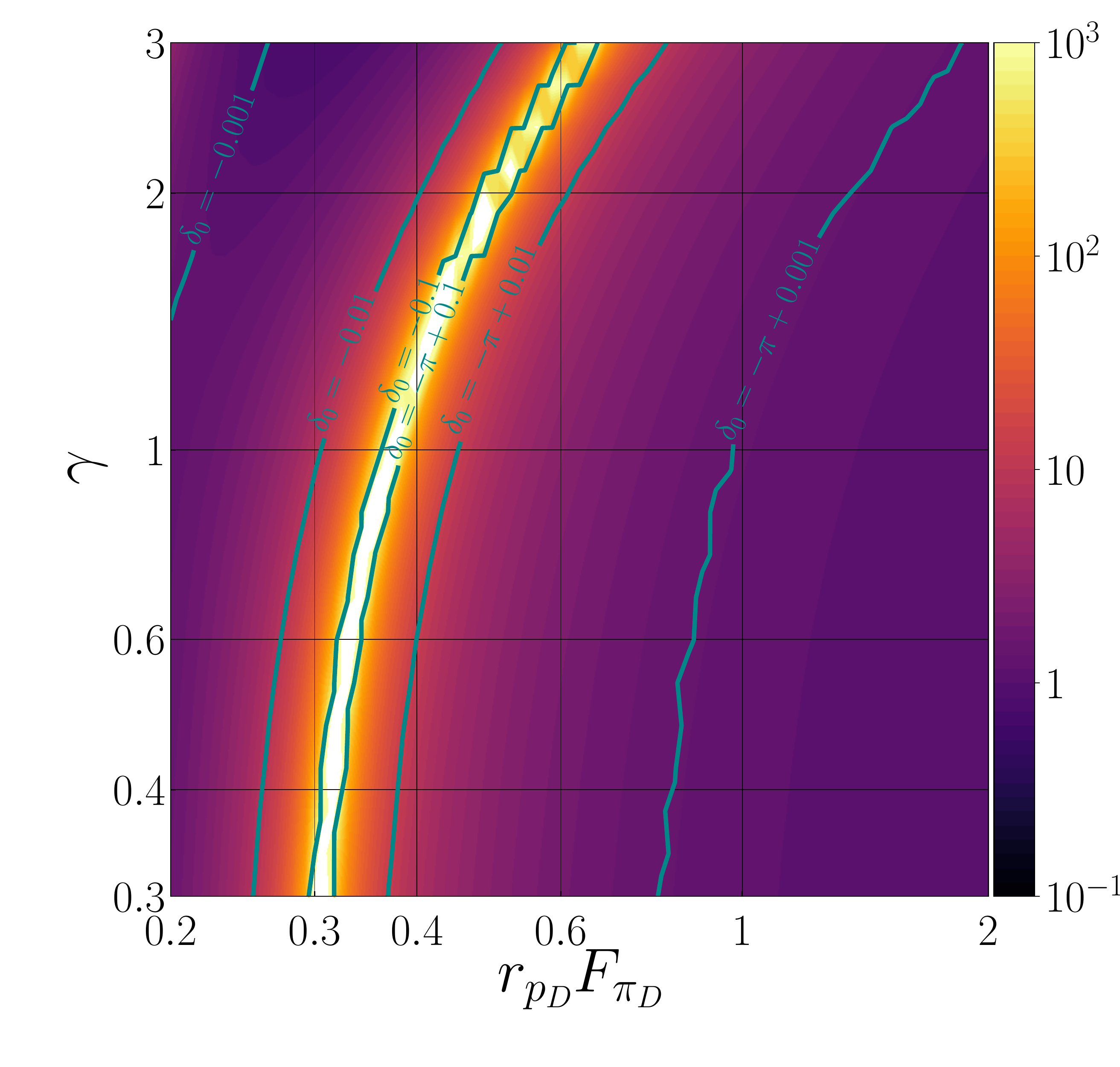}
\includegraphics[width=0.5\linewidth]{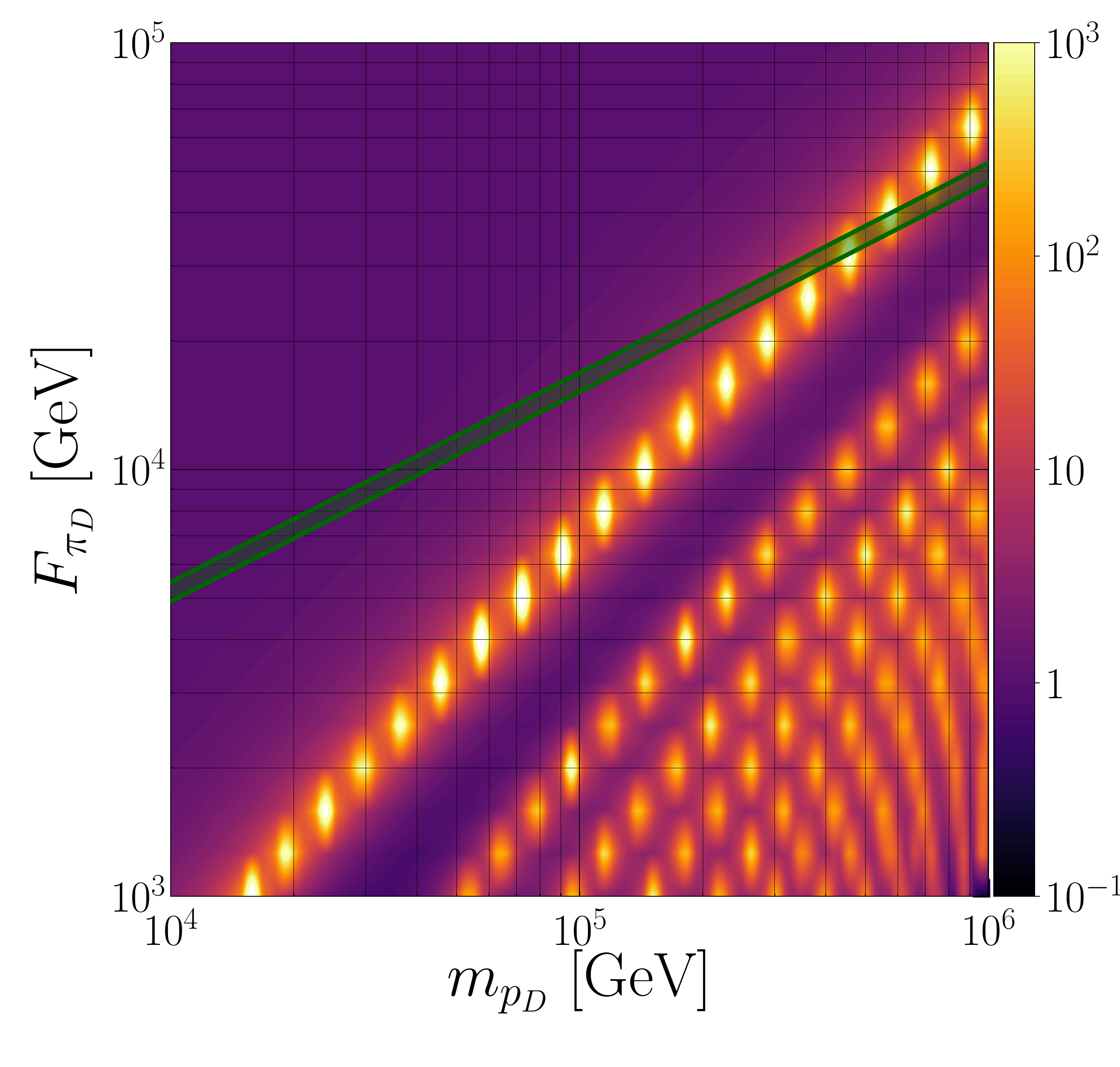}}
\caption{Left: Sommerfeld enhancement of $p_D \bar{p}_D$ annihilation (colors) and the low energy phase shift (contours labeled in blue) as a function of short-distance parameters $r_{p_D}$ (in units of $F_{\pi_D}^{-1}$) and $\gamma$, in the $S = 1$, $L = 0$ state, which couples to the $L = 2$, $J = 1$ state. We have fixed $m_{p_D}=200\,$TeV, $F_{\pi_D}=15\,$TeV, and the dark matter velocity $v_\text{rel} = 10^{-3}$. Right: Sommerfeld enhancement of $p_D \bar{p}_D$ annihilation as a function of the dark proton mass $m_{p_D}$ and dark pion decay constant $F_{\pi_D}$. The cutoff radius is fixed to be $0.4/F_{\pi_D}$ and $\gamma$ is chosen to be $1$. In the right figure, we also overlay the region of parameter space (green band) for which the relic density of dark matter is satisfied, per Fig.~\ref{fig:freezeout}. In both figures, $m_{\pi_D}=1\,$TeV and $g_A = 1.26$. }
\label{fig:SE:S=02}
\end{figure}

For $r\leq r_{p_D}$, the form of the potential is to be dictated by high-scale physics which cannot be calculated perturbatively in this model. In order to estimate the Sommerfeld enhancement factor, we parametrize the short-distance physics using a dimensionless parameter $\gamma$ and assume a box-shaped potential
\begin{eqnarray}\label{shortphysics}
\widetilde V_{C, T}(r) = \gamma V_{C, T} (r_{p_D}) , \hspace{1cm} ({\rm for} \ 0\leq r\leq r_{p_D}).
\end{eqnarray}
The boundary conditions at the origin are $R_{0k}(r\to 0) \to a$, 
$R_{0k}'(r\to0) \to 0$, 
$R_{2k}(r\to 0) \to b r^2$, 
$R_{2k}'(r\to 0) \to 2 b r$.
We vary the two boundary parameters $a$ and $b$ so that $R_{0k}$ and $R_{2k}$ match the partial waves of the incoming state at the infinity (see Appendix~\ref{appBC}),
\begin{eqnarray}\label{boundinf}
R_{0k}(r\to\infty) = \frac{\sqrt{\pi}}{kr} \cos\left[ kr - \frac{\pi}{2} + \delta_0 \right] \ , \ \ \ \ \ 
R_{2k}(r\to\infty) = \frac{\sqrt{2\pi}}{kr} \cos\left[ kr - \frac{3\pi}{2} + \delta_2 \right] \ .
\end{eqnarray}

The numerical results for the Sommerfeld enhancement factor $S_{L=0}$ and the phase shift $\delta_0$ in this coupled channel, as function of the short-distance parameters $r_{p_D}$ and $\gamma$, are depicted in Fig.~\ref{fig:SE:S=02}(left). Here the long-distance parameters --- $g_A = 1.26$, $F_{\pi_D} = 15$ TeV, $m_{p_D} = 200$ TeV and $m_{\pi_D} = 1$ TeV --- are held fixed. 

From an effective theory point of view, the phase shift $\delta_0$, as a low energy observable, should be insensitive to the details of the short-distance physics~\cite{Lepage:1997cs}. This implies that, in the dark $SU(3)$ model, once the long-distance parameters are fixed, the short distance parameters $r_{p_D}$ and $\gamma$ are no longer independent; they must be introduced in a way that leaves $\delta_0$ invariant. In other words, the correct long-distance model requires one to ``follow'' the  blue contours in the figure, {\it i.e.}, $\gamma$ is a function of $r_{p_D}$. It is interesting to notice from the figure that the Sommerfeld enhancement factor $S_{L=0}$ is also roughly constant along the constant $\delta_0$ contours and the value of $\delta_0$ determines the value of $S_{L=0}$ (and vice-versa). In this case, $\delta_0$ and $S_{L=0}$ are sensitive to UV physics ({\it i.e.}, the actual size of the dark proton, $r_{p_D}$) in the same fashion.\footnote{The fact that $S_{L=0}$ changes hand-in-hand with $\delta_0$ is easy to understand when the Sommerfeld enhancement is significant, which happens when an $s$-wave bound state is ``squeezed out'' of the potential well and converted into a resonance state (by changing the shape of the potential). In this case, the $p_D \bar p_D$ annihilation takes place dominantly through this intermediate resonant state. The propagator of the intermediate state changes sign when the incoming state energy lies above or below the pole, causing a jump in the phase shift (from 0 to $\pi$).}

Fig.~\ref{fig:SE:S=02}(right) depicts $S_{L=0}$ as a function of the long-distance parameters, $m_{p_D}$ and $F_{\pi_D}$.
This time we fix the short-distance parameters to be $\gamma=1$ and $r_{p_D}=0.4/F_{\pi_D}$.
We find the result is quite insensitive to $m_{\pi_D}$ for $m_{\pi_D}\ll m_{p_D}$. The reason is that,
in this case, the dominant contribution to the potential energy in Eq.~(\ref{CoupledSchrodinger}) is the $1/r^3$ term in $V_T$, which is $m_{\pi_D}$-independent. As with the one-state case discussed above, the potential only depends on the quantity $g_A/F_{\pi_D}$, so one may expect that the behavior observed in Fig.~\ref{fig:SE:S=02}(right) to also be obtained by varying $g_A$ instead. This is not the case for the coupled channel as the Sommerfeld enhancement depends significantly on the cutoff radius $r_{p_D}$.

Phenomenology-wise, in view of the current indirect detection experimental sensitivities, discussed in section~\ref{sec:DMID}, we conclude that within the bright yellow regions of Figs.~\ref{fig:SE:S=0} and \ref{fig:SE:S=02} (with $S_{L=0}\gtrsim 100$), the dark matter annihilation rate is large enough to be constrained by the present data, while next-generation experiments will probe our model over a large range of parameters. 

Another promising place where non-perturbative effects are important  is during the formation of the CMB, where the dark matter velocity is extremely low. In this case, we expect that the Sommerfeld Enhancement factor will saturate for $v \lesssim m_{\pi_D}/m_{p_D}$. For the spin-singlet (uncoupled) channel, we can use the results of Ref.~\cite{Feng:2010zp} and find that the Sommerfeld enhancement is strongly peaked, in terms of our model, for $g_A^2 m_{\pi_D} m_{p_D}/(8\pi^3 F_{\pi_D}^2) \simeq k^2$, where $k$ is any integer. The CMB constraint would be relevant near these regions, excluding a part of parameter space. We expect a similar effect in the spin-triplet (coupled) channel annihilation. We leave a detailed calculation of this constraint to a later work.

In addition to the Sommerfeld enhancement, the dark matter annihilation rate may be further enhanced if $p_D, \bar p_D$ form a bound state before annihilating. In models with vector-mediated or (real) scalar-mediated dark forces, bound-state effects are known to have important implications for indirect detection~\cite{Hisano:2011dc,Kawasaki:2015yya,An:2016gad, An:2016kie, Asadi:2016ybp, Petraki:2016cnz, Cirelli:2016rnw, Mitridate:2017izz, Wise:2014ola}. A complete calculation of bound state formation in the pseudoscalar case is beyond the scope of this work; we leave it for a future study. 

%%%%%%%%%%%%%%%%%%%%%%%%%%%%%%%%%%%%%%%%%%%%%%%%%%%%%%%%%%%%%%%%%%%%%%

\setcounter{footnote}{0}
\setcounter{equation}{0}
\section{Concluding Remarks and Other Comments}
\label{sec:conclusions}

All concrete evidence for phenomena outside the SM -- neutrino masses and dark matter -- is consistent with the existence of new degrees of freedom that interact very weakly, if at all, with those in the SM. Here we propose that these new degrees of freedom organize themselves into a simple dark sector, a chiral non-abelian gauge theory -- $SU(3)\times SU(2)$ with minimal, nontrivial fermion content. Similar to the SM, the gauge symmetry is spontaneously broken to $SU(3)$, which confines at low energies. Again similar to the SM, at the renormalizable level, the dark sector contains massless fermions -- dark leptons -- and stable massive particles -- dark protons. The stability of the dark proton is guaranteed by an accidental dark baryon number symmetry. We explore the possibility that the dark leptons play the role of right-handed neutrinos via neutrino-portal interactions and the possibility that the dark protons are the dark matter. We find that dark protons with masses between roughly 10--100~TeV satisfy all current cosmological and astrophysical observations concerning dark matter even if dark protons are a symmetric thermal relic, {\it i.e.}, even if there is no primordial dark baryon asymmetry. The dark leptons play the role of right-handed neutrinos and allow simple realizations of both the Type-I seesaw mechanism or the possibility that neutrinos are Dirac fermions. In the latter case, one naturally understands why neutrino masses are parametrically different from charged-fermion masses and predict the lightest neutrino to be massless. We highlight that, since our manifestation of the dark sector does not contain a $U(1)$ subgroup, there is no kinetic-mixing portal between the SM and the dark sectors. Many of the results highlighted here are a consequence of this fact.  

Since the new ``neutrino'' and ``dark matter'' degrees of freedom interact with one another, these two new physics phenomena are closely intertwined. Dark leptons play a nontrivial role in early universe cosmology and we find that cosmic surveys constrain these new degrees of freedom to be very light (either their Majorana masses are under 10~eV or the neutrinos are Dirac fermions) or relatively heavy (Majorana masses above 500~MeV). On the other hand, indirect searches for dark matter involve, decisively, dark matter annihilations into dark leptons. These, in turn, may lead to observable signatures at high-energy neutrino and gamma ray observatories. 

Throughout, we postulated the existence of a dark Higgs doublet $H_D$ that spontaneously breaks the $SU(2)$ gauge symmetry in the dark sector.  Instead, one could render the dark sector particle content more minimal by removing $H_D$ while still attaining many of the results discussed in this manuscript -- the dark quark bilinear $Q_D u_D^c$ could play the role of $H_D$. If there is no $H_D$ field, there are no renormalizable interactions between the SM and the dark sector. However, dimension-six operators, such as $Q_D \slashed{D} Q_D (H^\dagger H)$ (or even higher-dimension operators such as $(Q_D u_D^c)(\bar Q_D \bar u_D^c)(H^\dagger H)$) could serve the role of the Higgs portal and could be responsible for thermally equilibrating the SM and the dark sectors in the early universe. On the other hand, the dark SU(2) gauge symmetry is spontaneously broken together with dark chiral symmetry when the dark $SU(3)$ confines. Note that, absent $H_D$, the dark quarks are massless and dark chiral symmetry is exact. After symmetry breaking, the dark pions, here, are would-be goldstone bosons, ``eaten'' to become the longitudinal components of the $X$ gauge bosons. In this case, the dark proton could still serve as a symmetric thermal dark matter candidate\footnote{Because the dark sector $SU(2)$ symmetry breaking scale and the dark proton mass are of the same order, and freeze-out of the dark matter occurs at $T \sim m_{p_D}/30$, the dark sphaleron effects that could mix dark baryons and leptons are expected to be suppressed.}. It dominantly annihilates into longitudinal $X$ bosons (dark pions) as well as the $\eta'_D$ meson. Annihilations into transverse $X$ bosons are subdominant as long as the dark $SU(2)$ gauge coupling $g_2$ is perturbative. Naively, the dark proton mass should still be around 10--100\,TeV if it is to make up all of the dark matter. In this case, $\eta_D'$ decays into a pair of $X$ bosons, while the $X$ bosons mainly decay into pairs of dark leptons. As far as the neutrino sector is concerned, Majorana mass terms (in the context of Eq.~(\ref{eq:dim5})) for the left-handed antineutrinos take the form $(L_D Q_D u_D^c)(L_D Q_D u_D^c)$, $(\bar Q_D d_D L_D)(L_D Q_D u_D^c)$, $(\bar Q_D d_D L_D)(\bar Q_D d_D L_D)$ -- dimension-nine operators -- and the Dirac neutrino mass terms take the form $(L_D Q_D u_D^c)(LH)$, $(\bar Q_D d_D L_D)(LH)$ -- dimension 7 operators. In this scenario, it is possible to contemplate a connection between the dark strong interaction scale and the origin of the observed neutrino masses.

Before closing, we would also like to comment on the twin-Higgs models~\cite{Chacko:2005pe}, in particular the ``fraternal'' versions~\cite{Craig:2015xla, Craig:2015pha}, designed to address the hierarchy problem. These models consist of a hidden sector with the same gauge symmetry -- $SU(3)\times SU(2)$ -- explored here, even though the motivations that led us to it were quite distinct. There are, not surprisingly, several major differences between our proposals. First, we do not aim at addressing the hierarchy problem and focused only on the case where the dark quark Yukawa couplings are small enough so that the dark mesons and baryons are the lightest dark hadron states. The dark glueball states, for example, are heavy. At the same time, we have the freedom to consider dark $SU(3)$ confinement scales much higher than the electroweak symmetry breaking scale and were able to identify that dark protons with masses above 10~TeV can play the role of the dark matter in the absence of a primordial dark baryon asymmetry. Second, we never introduce dark $SU(2)$ singlet states (dark right-handed ``charged'' leptons). These are gauge singlets and (a) violate our minimalist aspirations, and (b) if present, are allowed Majorana masses completely divorced from all SM or dark sector mass scales. As a result, in our discussions, the $SU(2)$ doublet dark leptons are massless at the renormalizable level and it is possible to explore the possibility that these play the role of right-handed neutrinos in the context of explaining the origin of (Dirac or Majorana) neutrino masses. In our scenario, dark leptons can also serve as the portal for the indirect detection of dark matter.

\section*{Acknowledgments}
We are happy to thank Csaba Csaki and Yue Zhao for useful conversations.
This work is supported in part by the DOE grant \#DE-SC0010143.

\appendix
\section{Sterile Neutrino Lifetime}\label{sec:SterileNuAppendix}
\setcounter{equation}{0}

Here, we list the expressions for the partial widths that we used to determine the sterile neutrino lifetimes in Fig.~\ref{fig:nulife}. Most of these results come from Ref.~\cite{Gorbunov:2007ak}, the exception being the width for ${\nu_D} \to \nu_\alpha \gamma$, which we have taken from Ref.~\cite{Pal:1981rm}.
\begin{eqnarray}
\Gamma \left( {\nu_D} \to \sum_{\beta=e, \, \mu, \, \tau} \nu_\alpha \nu_\beta \overline{\nu_\beta} \right) & = & \frac{G_F^2 m_{\nu_D}^5}{192 \pi^3} |U_\alpha|^2, \\
\Gamma \left( {\nu_D} \to \nu_\alpha \gamma \right) & = & \frac{9  \alpha_{EM} G_F^2 m_{\nu_D}^5}{512 \pi^2} |U_\alpha|^2, \\
\Gamma \left( {\nu_D} \to \pi^0 \nu_\alpha \right) & = & \frac{G_F^2 f_\pi^2 m_{\nu_D}^3}{32 \pi} |U_\alpha|^2 \cdot \left(1 - \frac{m_{\pi^0}^2}{m_{\nu_D}^2} \right)^2,  \\
\Gamma \left( {\nu_D} \to \pi^+ \ell_\alpha^- \right) & = & \frac{G_F^2 f_\pi^2 m_{\nu_D}^3 |V_{ud}|^2}{16 \pi} |U_\alpha|^2 \cdot \left( \left(1-\frac{m_{\ell_\alpha}^2}{m_{\nu_D}^2} \right)^2 - \frac{m_{\pi^+}^2}{m_{\nu_D}^2} \left(1+\frac{m_{\ell_\alpha}^2}{m_{\nu_D}^2} \right) \right)^2 \nn \\
& & \times \sqrt{ \left( 1- \frac{(m_{\pi^+}-m_{\ell_\alpha})^2}{m_{\nu_D}^2} \right) \left( 1- \frac{(m_{\pi^+}+m_{\ell_\alpha})^2}{m_{\nu_D}^2} \right) }, \quad \left( \alpha = e, \, \mu \right) \\
\Gamma \left({\nu_D} \to \ell_\alpha^- \ell^+_\beta \nu_\beta  \right) & = &  \frac{G_F^2 m_{\nu_D}^5}{192 \pi^3} |U_\alpha|^2 \cdot \left(1 - 8x_\ell^2 + 8 x_\ell^6 - x_\ell^8 -12 x_\ell^4 \log x_\ell^4 \right), \nn \\
& & \qquad \qquad \qquad \qquad \qquad \qquad \left(\alpha \neq \beta; \, x_\ell \equiv \frac{\max\left[m_{\ell_\alpha}, m_{\ell_\beta} \right]}{m_{\nu_D}} \right) \\
\Gamma \left({\nu_D} \to \nu_\alpha \ell_\beta^- \ell^+_\beta \right) & = &  \frac{G_F^2 m_{\nu_D}^5}{192 \pi^3} |U_\alpha|^2 \cdot \Big[ \Big(C_1 (1-\delta_{\alpha\beta}) + C_3 \delta_{\alpha\beta} \Big) \times  \nn \\
& & \left. \Big( \left(1- 14y_\ell^2 - 2 y_\ell^4  - 12 y_\ell^6 \right) \sqrt{1-4 y_\ell^2}+ 12 y_\ell^4 (y_\ell^4 -1) L \Big)  \right. \nn \\
& & + 4 \Big(C_2 (1-\delta_{\alpha\beta}) + C_4 \delta_{\alpha\beta} \Big) \times \Big(y_\ell^2 \left(2+10y_\ell^2-12y_\ell^4 \right) \sqrt{1-4 y_\ell^2} \nn \\
& & + 6 y_\ell^4 \left(1 - 2y_\ell^2 + 2 y_\ell^4 \right) L \Big) \Big], \quad \qquad \qquad \qquad \qquad \left( y_\ell \equiv \frac{m_{\ell_\beta}}{m_{\nu_D}} \right)
\end{eqnarray}
where $\alpha,\beta=e,\mu,\tau$, $G_F$ is the SM Fermi constant, $\alpha_{EM}$ is the fine-structure constant of electromagnetism, $f_\pi$ is the SM pion decay constant, $V_{ud}$ is the appropriate element of the CKM matrix, $m_{\pi^0}$ ($m_{\pi^+}$) is the neutral (charged) SM pion mass, $m_{\ell_\alpha}$ is the mass of $\ell_{\alpha}$,
\begin{eqnarray}
C_1 = \frac{1}{4} \left(1-4 \sin^2 \theta_w + 8 \sin^4 \theta_w \right), & & C_2 = \frac{1}{2} \sin^2 \theta_w \left(2 \sin^2 \theta_w - 1 \right), \nn \\
C_3 = \frac{1}{4} \left(1+4 \sin^2 \theta_w + 8 \sin^4 \theta_w \right), & & C_4 = \frac{1}{2} \sin^2 \theta_w \left(2 \sin^2 \theta_w + 1 \right), \nn
\end{eqnarray}
where $\theta_w$ is the Weinberg angle, and
\begin{equation}
L = \log \left[ \frac{1-3y_\ell^2 - \left(1- y_\ell^2 \right) \sqrt{1-4y_\ell^2}}{y_\ell^2 \left(1+\sqrt{1-4y_\ell^2} \right)} \right].
\end{equation}
$U_{\alpha}=U_{\alpha4}$ or $U_{\alpha 5}$, depending on whether $\nu_D$ is the fourth or fifth neutrino mass eigenstate (as in the text, the mostly-sterile neutrinos $\nu_D$ are associated to $\nu_4$ and $\nu_5$) and the mostly-active neutrino mass eigenstates are treated as if they were massless. See Refs.~\cite{Pal:1981rm,Gorbunov:2007ak} for more details. The total decay width is given by
\begin{align}
\Gamma_{\rm total} & = 2 \times \left[ \sum_{\alpha=e, \, \mu, \, \tau} \Gamma \left( {\nu_D} \to \sum_{\beta=e, \, \mu, \, \tau} \nu_\alpha \nu_\beta \overline{\nu_\beta} \right) + \sum_{\alpha=e, \, \mu, \, \tau} \Gamma \left( {\nu_D} \to \nu_\alpha \gamma \right) + \sum_{\alpha=e, \, \mu, \, \tau} \Gamma \left( {\nu_D} \to \pi^0 \nu_\alpha \right) \right. \nn \\
& \left. + \sum_{\alpha=e, \, \mu} \Gamma \left( {\nu_D} \to \pi^+ \ell_\alpha^- \right) + \sum_{\alpha=e, \, \mu} \sum_{\beta=e, \, \mu} \Gamma \left({\nu_D} \to \ell_\alpha^- \ell^+_\beta \nu_\beta \right) + \sum_{\alpha=e, \, \mu, \, \tau} \sum_{\beta=e, \, \mu} \Gamma \left({\nu_D} \to \nu_\alpha \ell_\beta^- \ell^+_\beta \right) \right].
\end{align}
The overall factor of 2 accounts for the fact that the heavy, sterile neutrinos, because they are Majorana fermions, can decay both into the states we have explicitly listed as well as their $CP$ conjugates.

\begin{figure}[t]
\centerline{\includegraphics[scale=0.5]{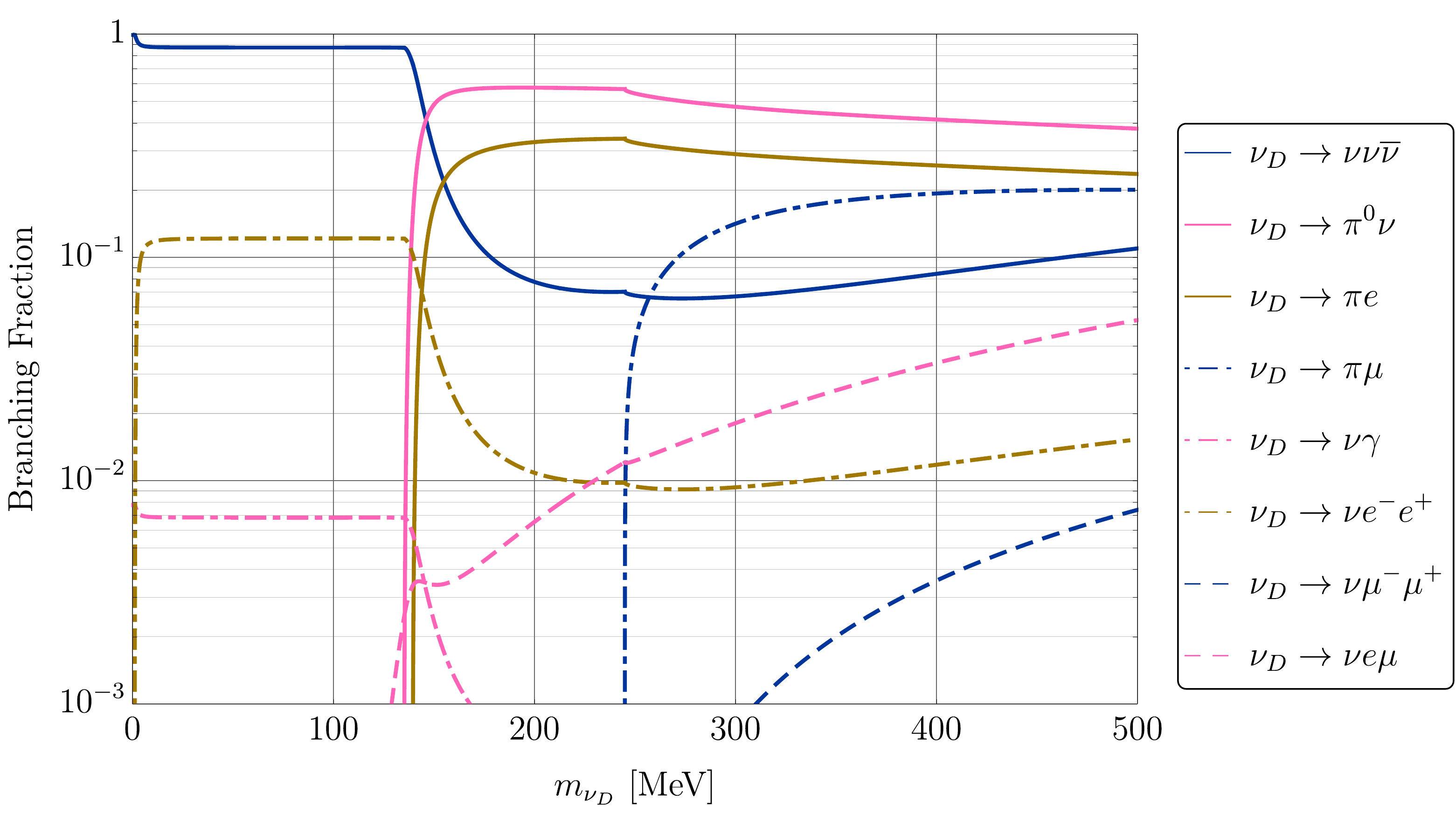}}
\caption{Branching fractions for the decay of a heavy, mostly-sterile neutrino as a function of its mass, $m_{\nu_D}$, calculated using the expressions for the width given in the text. For masses below $m_{\nu_D} \lesssim 500$ MeV, the following final states are relevant: $\nu \nu \overline{\nu}$ (solid blue), $\pi^0 \nu$ (solid pink), $\pi e$ (solid gold), $\pi \mu$ (dot-dashed blue), $\nu \gamma$ (dot-dashed pink), $\nu e^- e^+$ (dot-dashed gold), $\nu \mu^- \mu^+$ (long-dashed blue) and $\nu e \mu$ (long-dashed pink). The different curves apply simultaneously to the different $CP$-conjugated final state, e.g., the $\pi e$ curve includes decays to $\pi^- e^+$ as well as to $\pi^+ e^-$. The mostly-active neutrino flavors have been summed over.}\label{fig:nu_BRs}
\end{figure}

Fig.~\ref{fig:nu_BRs} depicts the branching fractions for the decay of the heavy, sterile neutrino, in order to contextualize the expressions above. Above $m_{\nu_D} \sim 500$ MeV, kaons become kinematically accessible and the number of potential final states becomes large; we do not show this region of parameter space, to avoid clutter. In the figure, we assumed all PMNS factors $|U_\alpha|^2$ equal to a common value, $|U|^2$, as portrayed in section~\ref{sec:neutrino}; the branching fractions thus do not depend on $|U|^2$.

\section{Coupled Eigenvalue Equations}\label{appSch}

Here we provide more details concerning the derivation of Eq.~(\ref{CoupledSchrodinger}).
The scattering state wavefunction in Eq.~(\ref{wave}) satisfies 
\begin{eqnarray}
\hat H  \Psi_{\vec{k}} (\vec{r}) = \left[\frac{\vec{p}^2}{2\mu} + V(r) \right]\Psi_{\vec{k}} (\vec{r}) = \frac{k^2}{2\mu} \Psi_{\vec{k}} (\vec{r}) \ ,
\end{eqnarray}
where the potential $V(r)$ is given by Eq.~(\ref{eq:Vpot}). In order to translate this eigenvalue equation into equations for $R_{0k}$ and $R_{2k}$, we choose the following representation for the
spin operator in $V(r)$ and the spin states in Eq.~(\ref{eq:CG}), for $S=1$,
\begin{eqnarray}
&&\hat S_x = \begin{pmatrix}
0 & 0 & 0\\
0 & 0 & -i \\
0 & i & 0
\end{pmatrix}, \ \ \ \hat S_y = \begin{pmatrix}
0 & 0 & i\\
0 & 0 & 0 \\
-i & 0 & 0
\end{pmatrix}, \ \ \ \hat S_z = \begin{pmatrix}
0 & -i & 0\\
i & 0 & 0 \\
0 & 0 & 0
\end{pmatrix} \ , \nonumber \\
&&
|SM_S\rangle =|11\rangle = - \frac{1}{\sqrt2} \begin{pmatrix}
1\\
i\\
0
\end{pmatrix}, \ \ \ |1-\!\!1\rangle = \frac{1}{\sqrt2} \begin{pmatrix}
1\\
-i\\
0
\end{pmatrix}, \ \ \ |10\rangle = \begin{pmatrix}
0\\
0\\
1
\end{pmatrix} \ .
\end{eqnarray}
With this representation, we project $\Psi_{\vec{k}} (\vec{r})$ onto the $|SM_S\rangle =|10\rangle$ subspace, yielding
\begin{eqnarray}
\left\langle 10|V(r)|\Psi_{\vec{k}} (\vec{r})\right\rangle \!\!&=&\!\! A V_C(r) \left[ R_{0k}(r) Y_{00}(\hat r) - \sqrt{\frac{2}{5}} R_{2k}(r) Y_{20}(\hat r) \right] \nonumber \\
\!\!&-&\!\! A V_T(r) \left[ -\sqrt{\frac{8}{5}} R_{2k}(r) Y_{20}(\hat r) - \sqrt{8} R_{2k}(r) Y_{00}(\hat r) + \frac{4}{\sqrt{5}} R_{0k}(r) Y_{20}(r) \right] + \cdots \ ,
\end{eqnarray}
where we defined $A = {g_A^2 m_{\pi_D}^2}/({48\pi F_{\pi_D}^2})$ and the ellipsis in the above equarion represent terms involving $Y_{LM}(\hat r)$ with $L>2$.
The coupled Schr\"odinger equations~(\ref{CoupledSchrodinger}) can be obtained by integrating the above equation with $\int d\Omega_{\hat r} Y^*_{00, 20}(\hat r)$.

\section{Boundary Conditions at Infinity}\label{appBC}

At infinity, a plane wave made of $p_D$ and $\bar p_D$ (including the spin degrees of freedom) can be written as
\begin{eqnarray}
&&\Psi_{\vec{k}} (r\to\infty) = \frac{1}{2} e^{i \vec{k} \cdot \vec{r}} \sum_{S=0}^1 \sum_{M_S=-S}^S | S M_S \rangle + | f (\hat r, S)\rangle \frac{e^{i k r}}{r}, \nonumber \\
&&= \frac{1}{2} \sum_{L M_L} \frac{4\pi i^L e^{i \delta_L}}{kr}\cos \left[ kr - \frac{\pi}{2} (L+1) + \delta_L \right] Y_{LM_L}^*(\hat k) Y_{LM_L}(\hat r) \sum_{S M_S} | S M_S \rangle, \nonumber \\
&&= \frac{1}{2} \sum_{J M_J} \sum_{L M_L}\sum_{S M_S} \frac{4\pi i^L e^{i \delta_L}}{kr}\cos \left[ kr - \frac{\pi}{2} (L+1) + \delta_L \right] Y_{LM_L}^*(\hat k) \langle \vec{r} | LSJM_J \rangle \langle JM_J | L M_L S M_S\rangle, \nonumber \\
&&\supset  \frac{1}{2}\sum_{L M_L} \frac{4\pi i^L e^{i \delta_L}}{kr}\cos \left[ kr - \frac{\pi}{2} (L+1) + \delta_L \right] Y_{LM_L}^*(\hat k) \langle \vec{r} |L110 \rangle \langle 10| L M_L 1 -\!\!M_L \rangle, \nonumber \\
&&\supset \frac{1}{2} \frac{\sqrt{4\pi} e^{i \delta_0}}{kr} \cos\left[ kr - \frac{\pi}{2} + \delta_0 \right] + \frac{1}{2} \frac{\sqrt{8\pi} e^{i \delta_2}}{kr} \cos\left[ kr - \frac{3\pi}{2} + \delta_2 \right] \ ,
\end{eqnarray}
where in the third line we have inserted a complete basis $\sum_{JM_J} | J M_J\rangle \langle J M_J |=1$, in the fourth line we have chosen to focus on the $J=S=1, M_J=0$ subspace, and in the last line we have chosen $\vec{k}$ to be along the $\hat z$ axis which forces $M_L=0$. Comparing with Eq.~(\ref{wave}), we obtain the boundary conditions Eq.~(\ref{boundinf}). 

\bibliographystyle{apsrev-title}
\bibliography{DM_bib}{}

\end{document}